\documentclass[aps, prb, twocolumn,superscriptaddress,amsmath,amssymb,reprint]{revtex4-2}
\usepackage{graphicx}% Include figure files
\usepackage{float} %设置图片浮动位置的宏包
\usepackage{subfigure} %插入多图时用子图显示的宏包
\usepackage{dcolumn}% Align table columns on decimal point
\usepackage{bm}% bold math
\usepackage[dvipsnames]{xcolor}
\usepackage{physics}
\usepackage{pifont}
\usepackage{comment}
\usepackage{makecell}
\usepackage{svg}
\usepackage{musixtex}
\usepackage[normalem]{ulem}
\begin{document}

\title{Giant and controllable nonlinear magneto-optical effects in two-dimensional magnets}

\author{Dezhao \surname{Wu}}
\affiliation{State Key Laboratory of Low Dimensional Quantum Physics and Department of Physics, Tsinghua University, Beijing, 100084, China}

\author{Meng \surname{Ye}}
\email{mye@gscaep.ac.cn}
\affiliation{Graduate School of China Academy of Engineering Physics, Beijing, 100193, China}

\author{Haowei \surname{Chen}}
\affiliation{State Key Laboratory of Low Dimensional Quantum Physics and Department of Physics, Tsinghua University, Beijing, 100084, China}

\author{Yong \surname{Xu}}
\email{yongxu@mail.tsinghua.edu.cn}
\affiliation{State Key Laboratory of Low Dimensional Quantum Physics and Department of Physics, Tsinghua University, Beijing, 100084, China}
\affiliation{Frontier Science Center for Quantum Information, Beijing, China}
\affiliation{RIKEN Center for Emergent Matter Science (CEMS), Wako, Saitama 351-0198, Japan}

\author{Wenhui \surname{Duan}}
\affiliation{State Key Laboratory of Low Dimensional Quantum Physics and Department of Physics, Tsinghua University, Beijing, 100084, China}
\affiliation{Frontier Science Center for Quantum Information, Beijing, China}
\affiliation{Institute for Advanced Study, Tsinghua University, Beijing 100084, China}

% \date{\today}
\begin{abstract}
The interplay of polarization and magnetism in materials with light
can create rich nonlinear magneto-optical (NLMO) effects, and the recent discovery of two-dimensional (2D) van der Waals magnets provides remarkable control over NLMO effects due to their superb tunability. 
Here, based on first-principles calculations, we reported giant NLMO effects in CrI$_3$-based 2D magnets, including a dramatic change of second-harmonics generation (SHG) polarization direction ($90^{\circ}$) and intensity (on/off switch) under magnetization reversal, and a 100\% SHG circular dichroism effect. 
We further revealed that these effects could not only be used to design ultra-thin multifunctional optical devices, but also to detect subtle magnetic orderings. 
Remarkably, we analytically derived conditions to achieve giant NLMO effects and propose general strategies to realize them in 2D magnets.
Our work not only uncovers a series of intriguing NLMO phenomena, but also paves the way for both fundamental research and device applications of ultra-thin NLMO materials.
\end{abstract}

\keywords{Suggested keywords}

\maketitle

%==========================
\section*{Introduction}
%===========================
%What is interesting and important.
The interaction of light with the polarization and magnetization in matters could create profound nonlinear magneto-optical (NLMO) phenomena, \textcolor{black}{such as the magnetization-induced second-harmonics generation (MSHG) and the photocurrent generation \cite{Wu_CrI3_2019,Yan-2019-NC-MIC,Qian-2020-npjCM-MIC}.} \textcolor{black}{Among them, NLMO effects related to second-harmonics generation (SHG) have distinctive advantages in both magnetization detection and light modulation.}
As an all-optical probe characterized by higher-rank tensors, \textcolor{black}{SHG-related} NLMO measurement is nondestructive with high spatial and temporal resolution \cite{Pisarev_hRMnO3_2000,Fiebig_rev_2005,Rasing_LPL_1997,Kirschner_Fe_1996,Kirilyuk_rev_2002}, and shows great promise to characterize structural and magnetic signals in 2D magnets \cite{2D-magnetic-probing_2017_nature,Mak_2019,Wu-2023-ACSnano}.
Moreover, the rotation angle of \textcolor{black}{SHG-related} NLMO effect is independent of sample thickness \cite{Koopmans_LPL_1995}, which is in clear contrast to the thickness-dependent linear magneto-optical Faraday angle, and therefore has the potential to be used in miniature devices.

\begin{figure*}
    \includegraphics[width=0.8\linewidth]{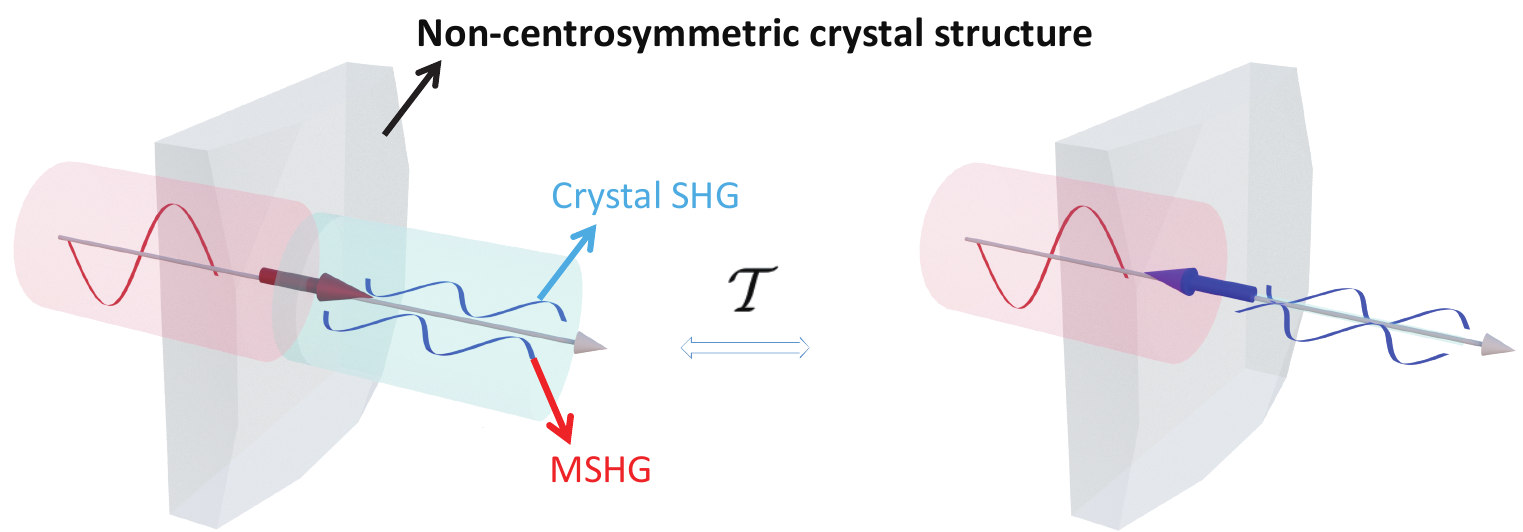}
    \caption{\textcolor{black}{Concept of NLMO effects. Magnets with non-centrosymmetric crystal structures exhibit both the crystal SHG and the MSHG due to the simultaneously breaking of $\mathcal{P}$, $\mathcal{T}$ and $\mathcal{PT}$ symmetries. The crystal SHG is unaffected by the magnetization direction reversal while the MSHG changes the sign. Thus the two terms can constructively (left panel) or destructively (right panel) interfere with each other for the positive and negative magnetization directions, respectively, which leads to NLMO effects. The red and blue arrows represent magnetic orders related by $\mathcal{T}$ operation. }  
     }
 \label{Fig1}
\end{figure*}

%Theory and Mechanism of NLMO
\textcolor{black}{The most significant SHG signal originates from electric dipole transitions \cite{Kirilyuk_rev_2002}, in which the breaking of the inversion ($\mathcal{P}$) symmetry is necessary. 
$\mathcal{P}$ symmetry can be broken by either crystal or magnetic structures, generating the corresponding crystal SHG and MSHG. 
These two types of SHG are described by an $i$-type tensor $\chi_{\rm N}$ which has even-parity under time-reversal ($\mathcal{T}$) operation and a $c$-type tensor $\chi_{\rm M}$ which is $\mathcal{T}$-odd \cite{Pershan_1963,Shen_1989,Birss_1964,Hubner_1989,shen1984principles}, respectively. 
Therefore, in non-centrosymmetric materials with $\mathcal{T}$ symmetry, only $\chi_{\rm N}$ survives, and in non-centrosymmetric magnetic materials with space-time inversion ($\mathcal{PT}$) symmetry, only $\chi_{\rm M}$ survives (Detailed derivations are in Supplementary Note 1A \cite{SI}).
When both types of SHG coexist, that is in materials with simultaneous breaking of $\mathcal{P}$, $\mathcal{T}$ and $\mathcal{PT}$ symmetries, a class of specific NLMO effects arises due to the interference of $\chi_{\rm N}$ and $\chi_{\rm M}$ under $\mathcal{T}$-operation as 
\begin{equation}
    \mathbf{P}(2\omega)\propto(\chi_{\rm N}\pm\chi_{\rm M})\mathbf{E}(\omega)\mathbf{E}(\omega)\,,
\end{equation}
where $\mathbf{E}(\omega)$ is the electric field of the incident light and $\mathbf{P}(2\omega)$ is the nonlinear polarization. 
The $\pm$ represents the influence of $\mathcal{T}$-operation. 
We denoted this interference effect of $\chi_{\rm N}$ and $\chi_{\rm M}$ as the NLMO effect in the rest of the context.
As illustrated in Fig. \ref{Fig1}, The interference between the two types of SHG can change the intensity of second harmonic (SH) light through $I(2\omega)\propto |\mathbf{P}(2\omega)|^2 \propto|\chi_{\rm N}\pm\chi_{\rm M}|^2$.
Thus, in magnets with comparable crystal SHG and MSHG, notable NLMO effects can emerge. }

%******************************************************************%
%******************************************************************%

%NLMO：experiments
However, experimentally observed NLMO effects in bulk materials are usually small 
%due to the incomparable $\chi_{\rm N}$ and $\chi_{\rm M}$ 
\cite{Kirschner_LPL_1991,Koopmans_LPL_1995,Rasing_LPL_1997,Kirilyuk_CPL_2000, Kirilyuk_LPL_2001,Train_2009,Tokura_FeGaO3_2004,Toyoda-2023-prm}, which seriously hinders the relevant research and applications of these effects. 
Additionally, a more profound theoretical understanding to explain and control the magnitude of NLMO effects is also absent. 
In contrast to bulk materials, recent experiments reveal many unusual SHG responses in two-dimensional (2D) materials, such as the large and tunable $i$-type SHG in MoS$_2$ \cite{MoS2_2013_nanoletters,MoS2_2014_Science}, NbOI$_2$/Cl$_2$ layers \cite{NbOI2_2021_nature-photonics,NbOCl2_2023_nature} and twisted h-BN \cite{twisted-hBN_2013_nanoletters,twist-hBN_2021_SA}, the giant pure $c$-type SHG in $\mathcal{PT}$-symmetric ultrathin CrI$_3$ \cite{Wu_CrI3_2019}, MnPS$_3$ \cite{MnPS3_2020_prl} and CrSBr \cite{CrSBr_2021_nanoletters}, and the anomalous SHG of uncertain origin in MnBi$_2$Te$_4$ thin films \cite{Xu_MBT_2022} and monolayer (ML) NiI$_2$ \cite{NiI2_2022_Nature}. 
These inspiring advances combined with the great tunability of atomically-thin materials made 2D materials an excellent platform for both fundamental research and design of miniature devices based on NLMO effects.

%Summary of our work
In this work, based on the computational method developed in Refs. \cite{Wang_NLO_Wannier,Chen_2022} and symmetry analysis, we investigated the NLMO effects of representative 2D magnets possessing both $\chi_{\rm N}$ and $\chi_{\rm M}$, including trilayer ABA stacking CrI$_3$, ML Janus Cr$_2$I$_3$Br$_3$ and ML H-VSe$_2$. 
We analytically derived and numerically calculated the NLMO angle for linearly-polarized incident light and the NLMO intensity asymmetry for circularly-polarized incident light at different frequencies.
Remarkably, we uncovered giant NLMO effects due to the maximal interference between comparable $\chi_{\rm N}$ and $\chi_{\rm M}$ in CrI$_3$-based 2D magnets, including a nearly 90$^\circ$ polarization rotation or an on/off switching of certain light helicity of SH light upon magnetization reversal, and a maximal SHG circular dichroism (SHG-CD) effect within fixed magnetization.
We also found that these NLMO effects are extremely sensitive to subtle changes in complex magnetic orders.
Moreover, we revealed the influence of interlayer interaction, spin-orbit coupling (SOC), and synergistic effect of stacking and magnetic orders to NLMO effects, and proposed strategies to achieve these giant NLMO effects in more 2D materials.
Our work not only uncovered a series of giant NLMO effects and corresponding candidate materials, but also provides strategies to manipulate these effects in other material systems, and further shed light on the application of NLMO effects in subtle magnetic orders detection and ultra-thin optical devices.

%=================================
\section*{Results and discussion} \label{Results and discussion}
%=================================

%--------------------------------------------------
 \subsection*{Structures and SHG of representative 2D magnets}
%--------------------------------------------------
%******************************************************************%
\begin{figure*}
    \includegraphics[width=1.0\linewidth]{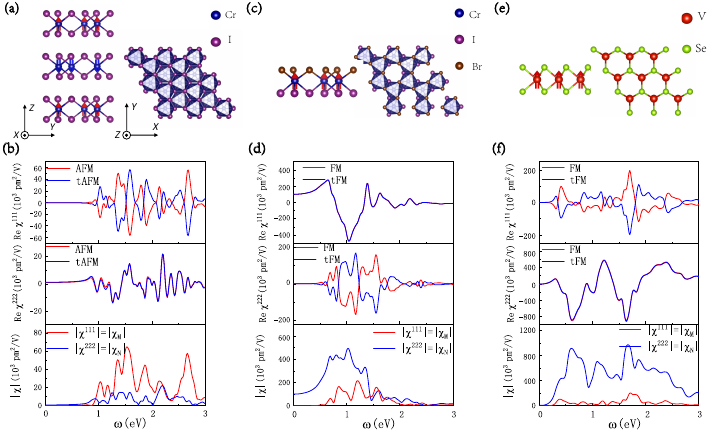}
    \caption{Atomic structures and SHG of the representative 2D magnets. 
    (a, c, e) Top and side views of ABA CrI$_3$ with AFM ($\uparrow$$\downarrow$$\uparrow$) magnetic order, ML Cr$_2$I$_3$Br$_3$ and ML H-VSe$_2$ with FM ($\uparrow$) magnetic order. The red/blue arrows represent the magnetic moment of each atom. (b, d, f) The real parts and the norm of SHG susceptibility tensor components. The red and blue lines represent results of  AFM ($\uparrow$$\downarrow$$\uparrow$)/FM ($\uparrow$) orders and their $\mathcal{T}$-related tAFM ($\downarrow$$\uparrow$$\downarrow$) /tFM ($\downarrow$) orders, respectively. }
 \label{Fig2}
\end{figure*}
%******************************************************************%
%~~~~~~~~~~~~~~~~~~~~~~~~~~~~~~~~~~~~~~~~~~~~~~~~~~~~~~~~~~~~~~~~~~%
\begin{table*} 
	\renewcommand{\arraystretch}{0.5}
	\centering
    \begin{tabular}{cccccc}
        \hline \hline 
         & \textcolor{black}{AB-AFM CrI$_3$} &  ABA-AFM CrI$_3$ & ML Cr$_2$I$_3$Br$_3$ & ML H-VSe$_2$ & ABA-MIX CrI$_3$ \\ \hline \
         Magnetic point group  &\textcolor{black}{$-3^{\prime}$}& $32^{\prime}$ & $3m^{\prime}$ & $-6m^\prime 2^\prime$ & $3$ \\ \hline 
         $\chi_{\rm N}$ &\textcolor{black}{NA}& 222=-211=-112 & 111=-122=-212 & 222=-211=-112 & \thead{111=-122=-212\\222=-211=-112}  \\ \hline 
         $\chi_{\rm M}$ &\textcolor{black}{\thead{111=-122=-212\\222=-211=-112}}& 111=-122=-212 & 222=-211=-112 & 111=-122=-212 & \thead{111=-122=-212\\222=-211=-112} \\ \hline \hline
    \end{tabular}
    \caption{Magnetic symmetries and symmetry properties of SHG tensor components of representative 2D magnets. $\chi_{\rm N}$ and $\chi_{\rm M}$ represent the $\mathcal{T}$-even and $\mathcal{T}$-odd SHG components. If a tensor component has both $\chi_{\rm N}$ and $\chi_{\rm M}$ parts, it means this tensor component does not have a definite parity under $\mathcal{T}$.} 
	\label{tab:sym}
\end{table*}
%~~~~~~~~~~~~~~~~~~~~~~~~~~~~~~~~~~~~~~~~~~~~~~~~~~~~~~~~~~~~~~~~~~%

NLMO effects exist in magnetic materials without $\mathcal{PT}$ symmetry and therefore, we consider two typical ferromagnetic (FM) semiconductors in the monolayer limit, CrI$_3$ and H-VSe$_2$ 
\cite{mcguire_CrI3_2015,sivadas_CrI3_2018,jiang_CrI3_2019,jang_CrI3_2019_microscopic,soriano_CrI3_2019,bonilla_VSe2_2018,wang_VSe2_2021}. 
ML VSe$_2$ is non-centrosymmetric with magnetic symmetry $\bar{6}m^\prime 2^\prime$ when the magnetization is along $\hat{z}$ axis. 
ML CrI$_3$ is centrosymmetric with \textcolor{black}{magnetic symmetry $\bar{3}m^{\prime}$ \cite{Yang_CrI3_2020} and AB stacking bilayer CrI$_3$ has $\mathcal{PT}$ symmetry as shown in Supplementary Figure 2 \cite{SI}, therefore the ML does not have SHG and the bilayer only has the MSHG $\chi_{\rm M}$.}
Therefore, we use the strategy of multi-layer stacking and element replacement to break ${\mathcal P}$ and ${\mathcal PT}$ to enable NLMO effects in CrI$_3$ related materials.
The simplest examples are ABA stacking trilayer CrI$_3$ with antiferromagnetic (AFM) interlayer coupling (denoted by ABA-AFM CrI$_3$) and ML Janus Cr$_2$I$_3$Br$_3$.
Figure \ref{Fig2}(a, c, e) shows the atomic and magnetic structures of ABA-AFM CrI$_3$, ML Cr$_2$I$_3$Br$_3$ and ML H-VSe$_2$ with their magnetic symmetries summarized in Tab. \ref{tab:sym}. Their band structures are shown in Supplementary Note 5 \cite{SI}.

%SHG susceptibilities 
Figure \ref{Fig2}(b, d, f) shows the influence of $\mathcal{T}$ operation to different SHG components $\chi^{abc}(2\omega;\omega,\omega)$ of the above mentioned materials. In the rest of the article, we use the shorthand notation $\chi^{abc}$ to represent $\chi^{abc}(2\omega;\omega,\omega)$, where $a,b,c$ are Cartesian directions. The two magnetic orders related by $\mathcal{T}$ symmetry are denoted as AFM/FM and tAFM/tFM, respectively. 
For SHG susceptibilities of trilayer CrI$_3$, we scissor the band gap to 1.5 eV and our results show good agreement with the previous work \cite{Yang_CrI3_2020} (\textcolor{black}{Supplementary Figure 3} \cite{SI}).
Due to the C$_{3z}$ symmetry in all three representative materials, each material has only two independent in-plane SHG components, that is $\chi^{111}=-\chi^{122}=-\chi^{212}$ and $\chi^{222}=-\chi^{211}=-\chi^{112}$, where the subscripts 1 and 2 denote the Cartesian direction $\hat{x}$ and $\hat{y}$.
Due to the presence of $m^\prime$ or $2^{\prime}$ symmetry in all three materials, one of the SHG component is $\mathcal{T}$-odd while the other is $\mathcal{T}$-even, as shown in the upper and middle panels of Fig. \ref{Fig2}(b, d, f). 
Generally, the even and odd quantities can coexist in the same tensor component and in this case, $\chi_{\rm N} = (\chi + \mathcal{T}\chi)/2$ and $\chi_{\rm M} = (\chi - \mathcal{T}\chi)/2$, where $\mathcal{T}\chi$ is the SHG susceptibility of the time-reversal pair. The detailed expressions of $\chi_{\rm N}$ and $\chi_{\rm M}$ \textcolor{black}{, as well as their symmetry requirements} are summarized in Supplementary Note 1 \cite{SI}. 
The parity of each component under $\mathcal{T}$ is also summarized in Tab. \ref{tab:sym}.

%relative ratio of N and M
The bottom panel of Fig. \ref{Fig2}(b, d, f) clearly shows that the relative value of $|\chi_{\rm N}|$ and $|\chi_{\rm M}|$ are distinctive in the three materials. 
Although $|\chi_{\rm M}|$ exceeds $|\chi_{\rm N}|$ in ABA-AFM CrI$_3$ at a wide frequency range and the opposite is observed in ML Cr$_2$I$_3$Br$_3$, there are still several intersections of $|\chi_{\rm N}|$ and $|\chi_{\rm M}|$.
In contrast, $|\chi_{\rm N}|$ is much larger than $|\chi_{\rm M}|$ in ML H-VSe$_2$ without any intersections.

%--------------------------------------------------
 \subsection*{Giant NLMO effects of representative 2D magnets}
%--------------------------------------------------

%******************************************************************%
\begin{figure*}
    \includegraphics[width=\linewidth]{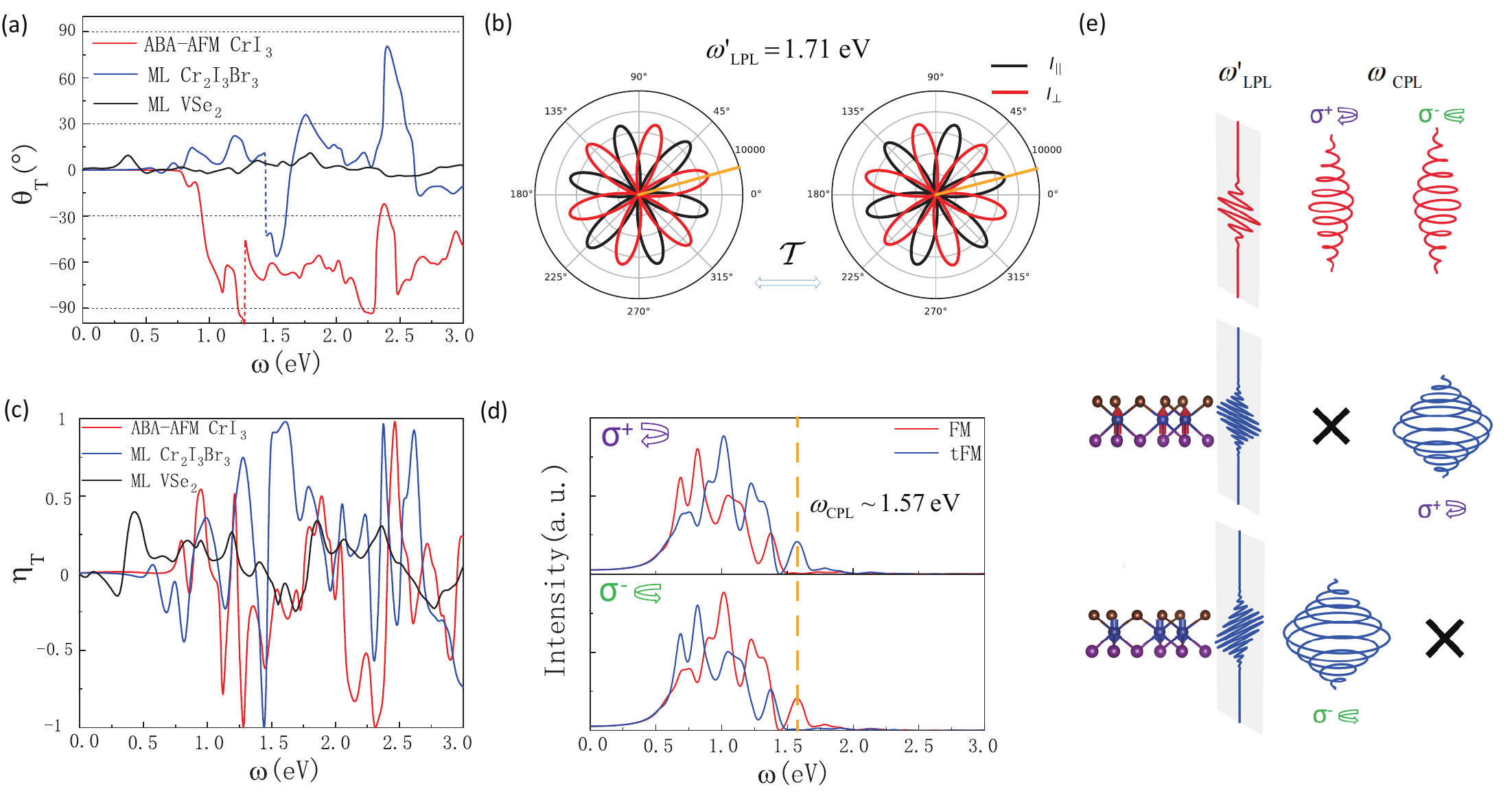} 
    \caption{NLMO effects in representative 2D magnets. (a) The NLMO angle (c) NLMO intensity asymmetry as a function of frequency. \textcolor{black}{The vertical dashed lines in the curves indicate the NLMO rotation is ill-defined at the corresponding frequencies and the polarization resolved SHG close to those frequencies in ABA-CrI$_3$ and ML Cr$_2$I$_3$Br$_3$ are shown in Supplementary Figure 5a and 7a \cite{SI}.}
    (b) Change of polarization-resolved SHG under $\mathcal{T}$ operation at a LPL characteristic frequency $\omega^{\prime}_{\rm LPL}=1.71$ eV of ML Cr$_2$I$_3$Br$_3$. The yellow line marks one of the LPL incident angle where the maximal polarization direction change of SHG happens under $\mathcal{T}$ operation.
    (d) CPL SHG intensity of FM and tFM states of Cr$_2$I$_3$Br$_3$. The upper and lower panels are for incident helicity $\sigma^{+}$ and $\sigma^{-}$. The difference between the red and blue curves in each panel reflects the NLMO intensity asymmetry. For a particular magnetic order, the difference between the upper and lower panels reflects the magnitude of SHG-CD.
    The dashed orange line indicates one of the CPL characteristic frequency $\omega_{\rm CPL}=1.57$ eV of ML Cr$_2$I$_3$Br$_3$.
    (e) Multi-degree controlling of NLMO in ML Cr$_2$I$_3$Br$_3$. 
    For linearly-polarized incident light at polarization angle $\theta=\pi/12+n\pi/6$ and frequency $\omega^{\prime}_{\rm LPL}$, ML Cr$_2$I$_3$Br$_3$ can act as an magneto-optical polarization switcher. For circularly-polarized incident light at frequency $\omega_{{\rm CPL}}$, ML Cr$_2$I$_3$Br$_3$ can act as a magneto-optical switch for circularly polarized SH light or an optical filter for circularly polarized SH light with a particular helicity.
    } \label{Fig3}
\end{figure*}
%******************************************************************%
The coexistence of $\chi_{\rm N}$ and $\chi_{\rm M}$ can induce a variety of NLMO effects. In the following, we calculated the SHG responses under the linearly and the circularly polarized light (LPL and CPL), with an emphasis on the role of different ratios of $\chi_{\rm N}$ and $\chi_{\rm M}$ to NLMO effects. 
Considering a normal incidence geometry where the material is in the $xy$-plane and the incident light propagates in the $-\hat{z}$ direction, the SHG polarization $\mathbf{P}(2\omega)$ of 2D materials has been given in Ref. \cite{Yang_CrI3_2020}. 
As the intensity of the emitted SH light $I(2\omega)$ is proportional to $|\mathbf{P}(2\omega)|^2$ and is commonly measured, we calculated $|\mathbf{P}(2\omega)|^2$ to represent $I(2\omega)$.

%SHG under LPL: general features without mentioning T symmetry
For LPL characterized by $\mathbf{E}=E(\cos\theta, \sin\theta)$, where $\theta$ denotes the angle between light polarization and the $x$-axis of the sample, we investigated the polarization-resolved SHG as a function of $\theta$.
The experiment measures the parallel ($||$) and perpendicular ($\bot$)  components of the SHG signal with respect to the direction of the incoming light polarization while the sample rotates with the angle $\theta$.
In the three representative materials, the SHG intensity at the parallel and perpendicular polarization directions can be written as \cite{Yang_CrI3_2020}
\begin{equation}
    \begin{aligned}
        I_{||}(\theta) & \propto | {\chi^{111}}  \cos{3\theta} - \chi^{222} \sin{3\theta} |^2 E^4 \,,\\
        I_{\bot}(\theta) & \propto  | {\chi^{111}}  \sin{3\theta} + \chi^{222} \cos{3\theta} |^2 E^4 .
    \end{aligned}\label{eq:intensity-L}
\end{equation}
%This equation works both for AFM and MIX CrI$_3$.
%General features: features without mentioning T
Both the parallel and perpendicular components exhibit sixfold sunflower-like patterns and differ only by a $\pm \pi/6$ angle, as the black and red lines shown in Fig. \ref{Fig3}(b).
In addition, the angle $\theta_{\rm m}$ corresponding to the maximum (minimum) of $I_{||}$ also corresponds to the minimum (maximum) of $I_{\bot}$.

%SHG under LPL: T-related features.
NLMO effects under LPL are reflected by comparing the polarization-resolved SHG patterns before and after $\mathcal{T}$ operation which results from the sign reverse of $\chi_{\rm M}$.
Under the symmetry of the three representative materials, this corresponds to a rotation of the polarization-resolved SHG pattern (Supplementary Figure 5b and 7b \cite{SI}), denoted as NLMO angle $\theta_{\mathcal{T}}$ (see details in Supplementary Note 2A \cite{SI}), with the expression 
%discussion on the expression
\begin{equation}
    \theta_{\mathcal{T}}= 2\theta_{\rm m}=-\frac{\tan^{-1} \left( \tan{2\delta} \cos{\Delta \varphi} \right)}{3} + \frac{n\pi}{3} \,,
    \label{eq:NLMOangle}
\end{equation}
where $n$ is an integer originated from the rotation symmetry of the pattern.
$\tan\delta = \frac{|\chi^{111}|}{|\chi^{222}|} \in (0, +\infty)$ is the magnitude ratio of $\chi^{111}$ and $\chi^{222}$, and $\Delta \varphi = -i\left( \ln{\frac{\chi^{111}}{|\chi^{111}|}} - \ln{\frac{\chi^{222}}{|\chi^{222}|}}\right) \in (-\pi, +\pi]$ is the phase difference between $\chi^{111}$ and $\chi^{222}$.
The range of $\delta$ is (0, $\pi/2$).
Eq. (\ref{eq:NLMOangle}) does not hold when $|\chi^{111}| = |\chi^{222}|$ and $\Delta\varphi = \pm \pi/2$. 
In this situation, the polarization-resolved SHG is isotropic according to Eq. (\ref{eq:intensity-L}), therefore it does not show any change under $\mathcal{T}$ operation.

%Discuss the nonlinear rotation angle
The maximum NLMO angle can be achieved is $\theta_{\mathcal{T}} = \pi/6 + n\pi/3$ with the condition
\begin{equation}
    \begin{cases}
        |\chi^{111}|= |\chi^{222}| \,,\\
        \Delta \varphi \neq \pm \pi/2 \,,
\end{cases}\label{eq:NLrotation}
\end{equation}
and we denoted the corresponding frequency as the LPL characteristic frequency $\omega_{\rm LPL}$.
At $\omega_{\rm LPL}$, the patterns of $I_{||}$ and $I_{\bot}$ are swapped exactly as shown in Fig. \ref{Fig3}(b), 
which results in a remarkable polarization change of the SH light.
%,especially for the incident light at $\theta_{\rm m}=\pi/12+n\pi/6$. 
If the magnitude of $I_{||}$ and $I_{\bot}$ differs greatly, the polarization of the SH light can rotate nearly $\pi/2$ under $\mathcal{T}$. 
The condition to achieve the $\pi/2$ polarization rotation at $\theta_{\rm m}=\pi/12+n\pi/6$ is
\begin{equation}
    \begin{cases}
        |\chi^{111}|= |\chi^{222}| \,,\\
        \Delta \varphi = 0 \, \rm or \, \pi \,,
    \end{cases}\label{eq:NLrotation'}
\end{equation}
and the corresponding frequency is denoted as $\omega^\prime_{\rm LPL}$, which is a subset of $\omega_{\rm LPL}$.
\textcolor{black}{This condition is also graphically shown in the complex plane in Supplementary Figure 1 \cite{SI}.}

%nonlinear rotation in the three materials
%We labeled the frequency satisfying Eq. \ref{eq:NLrotation} as LPL characteristic frequency $\omega_{LPL}$ and these are material-dependent characters. 
We numerically traced the $\omega$-dependence of NLMO angle  $\theta_{\mathcal{T}}$ in the three representative materials, as shown in Fig. \ref{Fig3}(a). 
As highlighted by the dashed horizontal line, $\theta_{\mathcal{T}}$ close to $\pi/6 + n\pi/3$ corresponds to the largest NLMO angle.
Remarkably, $\theta_{\mathcal{T}}$ is giant in ABA-AFM CrI$_3$ and ML Cr$_2$I$_3$Br$_3$ at several $\omega_{\rm LPL}$ frequencies. 
In contrast, $\theta_{\mathcal{T}}$ is always tiny in ML H-VSe$_2$. 
This distinction is consistent with their different ratio of $|\chi_{\rm N}|$ and $|\chi_{\rm M}|$ susceptibilities observed in Fig. \ref{Fig2}(b, d, f). 
Fig. \ref{Fig3}(b) shows the dramatic change of SHG light from perpendicular to parallel polarization direction for incident light at $\theta_m=\pi/12$ (indicated by the yellow line) under $\mathcal{T}$ in this atomically thin Cr$_2$I$_3$Br$_3$ at $\omega^{\prime}_{\rm LPL}=1.71$\,eV.

%SHG under CPL：general features without T
Next, we investigated the polarization and intensity of SH light under the illumination of the CPL.
For the CPL characterized by  $\mathbf{E} = E(1,\pm{i})$, where $\pm$ represents the left/right-handed ($\sigma^{\pm}$) CPL, the polarization of the SHG signal is 
\begin{equation}
    \mathbf{P}(2\omega) = 2E^2 (\chi^{111} \mp i\chi^{222}) (1,\mp{i})\,
    \label{eq:Polarization-C}
\end{equation}
according to Tab. \ref{tab:sym}. 
The SH light is also CPL with helicity opposite to the incident light. 
Furthermore, the intensity of the SH light depends on the helicity of the incident light, which is called the SHG-CD effect \cite{Hicks_SHGCD_1993,Hicks_SHGCD_1994,Cai_SHGCD_2014}.
The SHG-CD intensity asymmetry can be described by
\begin{equation}
    \eta_{\rm h}^\mathbf{M} = \frac{ I^{+} (\mathbf{M}) - I^{-} (\mathbf{M})}{I^{+} (\mathbf{M}) + I^{-} (\mathbf{M})} \,,
\end{equation}
where $I^{+}(\mathbf{M})$ is the SHG intensity generated by the magnetic order $\mathbf{M}$ when the helicity of the incident light is $\sigma^+$.

%SHG under CPL：features related to T
$\mathcal{T}$ operation can also change the intensity of the SH light with certain helicity. 
We can define the NLMO intensity asymmetry at certain incident light helicity $\sigma$ as
\begin{equation}
    \eta_{\mathcal{T}}^{\sigma} = \frac{I^{\sigma}(+\mathbf{M}) - I^{ \sigma}(-\mathbf{M})}{I^{\sigma}(+\mathbf{M}) + I^{ \sigma}(-\mathbf{M})}\,,
\end{equation}
where the $\pm \mathbf{M}$ states are $\mathcal{T}$-related, and $I^{\sigma}(+\mathbf{M})$ is the SHG intensity generated by the $+\mathbf{M}$ magnetic order.
The range of $\eta_{\mathcal{T}}^{\sigma}$ is [-1,1], and $\eta_{\mathcal{T}}^{\sigma}=\pm 1$ represents the situation in which an incident CPL with helicity $\sigma$ can only generate SH light in the $\pm \mathbf{M}$ magnetic state and are completely blocked in the opposite magnetic state. 
According to Eq. (\ref{eq:Polarization-C}), the reverse of helicity and the reverse of magnetization are equivalent, and therefore  
the change of intensity under $\mathcal{T}$ operation can also be reproduced by changing the helicity of CPL at a fixed magnetization, i.e., $\eta_{\mathcal T}^{+} = \eta_{\rm h}^{+\mathbf{M}}$.

%******************************************************************%
\begin{figure*}
    \includegraphics[width=\linewidth]{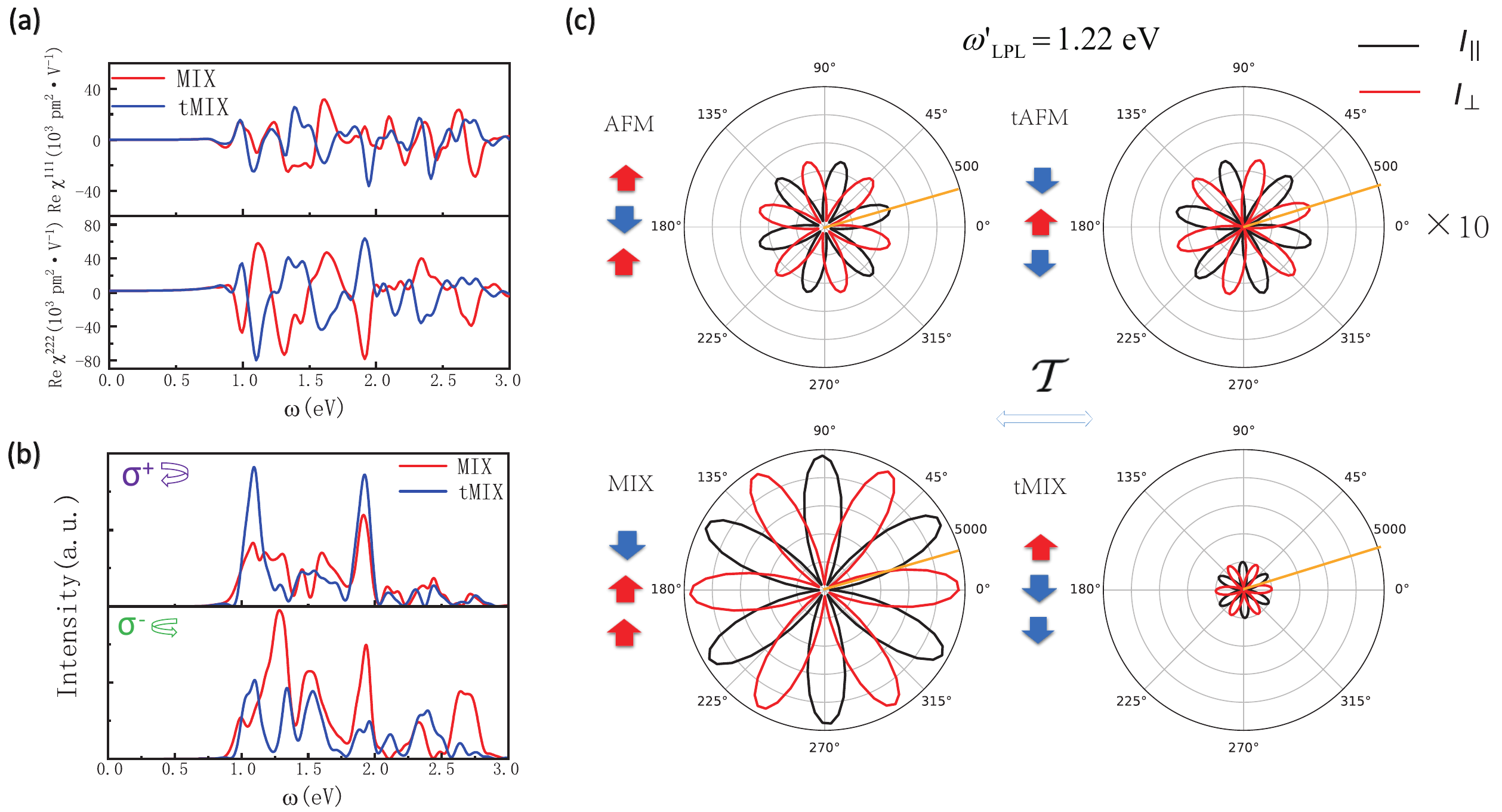}
    \caption{SHG susceptibility, CPL SHG intensity and polarization-resolved SHG for
    ABA-MIX CrI$_3$.
    (a) Real parts of SHG susceptibility tensor components of ABA-MIX CrI$_3$. Red and blue lines represent results for MIX($\uparrow$$\uparrow$$\downarrow$) and tMIX($\downarrow$$\downarrow$$\uparrow$) magnetic orders, respectively. 
    (b) SHG intensity of MIX and tMIX magnetic orders of ABA-MIX CrI$_3$ under CPL illumination. The upper and lower panels are for incident helicity $\sigma^+$ and $\sigma^{-}$. The difference between the red and blue curves in each panel reflects the $\eta_{\mathcal{T}}$. For a particular magnetic order, the difference between the upper and lower panels reflects the magnitude of SHG-CD. 
    (c) Polarization-resolved SHG of AFM, tAFM, MIX, and tMIX magnetic orders at one of the LPL frequency $\omega^\prime_{\rm LPL}$=1.22 eV of ABA-AFM CrI$_3$. The red/blue arrow represents the magnetic moment in each CrI$_3$ layer.
}\label{Fig4}
\end{figure*}
%******************************************************************%

%discussion on the expression
The expression of the NLMO intensity asymmetry in our case is (derivation is in Supplementary Note 2B \cite{SI}.)
\begin{equation}
    \eta_{\mathcal{T}}^{\pm}  = \mp\sin{2\delta} \sin{\Delta \varphi}\,.
\end{equation}
As long as $\Delta \varphi \neq n\pi$, the NLMO intensity asymmetry is always nonzero, and the intensity of the SH light is always magnetization-dependent. 
In general, the maximum $|\eta_\mathcal{T}^{\pm}|= 1$ requires  
\begin{equation}
    \begin{cases}
        |\chi^{111}| = |\chi^{222}| \,,\\
        \Delta \varphi = \pm \pi/2 \,,
\end{cases}\label{eq:CPLasymmetry}
\end{equation}
and we denoted the corresponding frequency as the CPL characteristic frequency $\omega_{\rm CPL}$. 
\textcolor{black}{This condition is also graphically shown in the complex plane in Supplementary Figure 1 \cite{SI}.}
It is worth noting that the $\omega_{\rm CPL}$ is just where the $\theta_{\mathcal{T}}$ for LPL is ill-defined.

We numerically traced the $\omega$-dependence of $\eta_{\mathcal{T}}^{+}(2\omega)$ in the three representative materials in Fig. \ref{Fig3}(c). 
Again, two CrI$_3$-based materials can achieve large $\eta_{\mathcal{T}}^{+}$ including $\pm 1$ at several frequencies while $|\eta_{\mathcal{T}}^{+}|$ is always below 0.4 in ML H-VSe$_2$. 
This contrast is another feature stemming from their distinct ratio of $|\chi_{\rm N}|$ and $|\chi_{\rm M}|$.
We take ML Cr$_2$I$_3$Br$_3$ as an example to illustrate the intensity change of SH light under $\mathcal{T}$ operation for $\sigma^{\pm}$ CPL in Fig. \ref{Fig3}(d). (The CPL SHG intensity of ABA-AFM CrI$_3$ and ML H-VSe$_2$ are shown in Supplementary Figure 4 and 8 \cite{SI}.) 
The vertical dash line in Fig. \ref{Fig3}(d) highlights one of the frequency $\omega_{\rm CPL}$ at which the intensity of SHG changes between a considerable value and zero under $\mathcal{T}$ operation. 
The upper ($\sigma^{+}$) and lower ($\sigma^{-}$) panels of Fig. \ref{Fig3}(d) looks exactly the same except the line colors are swapped, which reflects the equivalency of helicity and magnetization reversal.
%the relation $\eta_{\mathcal{T}}^{+} = -\eta_{\mathcal{T}}^{-}$. 
Other than that, the difference of the red (blue) line between the upper and lower panels in Fig. \ref{Fig3}(d) also indicates that this material exhibit a 100\% SHG-CD effect at $\omega_{\rm CPL}$.

%summary of applications
The giant NLMO effects discovered in CrI$_3$-based materials enabled them to be used as magnetic-field-controlled atomically thin optical devices, as illustrated in Fig. \ref{Fig3}(e). 
Utilizing incident light at frequencies $\omega^{\prime}_{\rm LPL}$, they can serve as optical polarization switchers for LPL as they can rotate the polarization of SH light by nearly $\pi/2$. 
At the incident frequency $\omega_{\rm CPL}$, they are ideal magneto-optical switches for CPL due to their large $\eta_{\mathcal{T}}$. 
In addition, even without an external magnetic field, due to their large SHG-CD effects, they can serve as optical filters for CPL with a particular helicity. 

\textcolor{black}{It is worth noting that the above discussed NLMO and SHG-CD effects are analyzed base on the nonlinear polarization $\mathbf{P}(2\omega)$ generated in materials and the intensity of radiation is estimated by $I(2\omega)\propto |\mathbf{P}(2\omega)|^2$, which is not specific for the reflected or the transmitted SH light. The exact value of reflected and transmitted SH light can be obtained by solving Maxwell equations with corresponding boundary conditions \cite{Chen-2022-npjCM-SHG_computation}. 
Although the SH light in reflection or transmission may have different intensity, the above discussed NLMO and SHG-CD effects persist.}

%******************************************************************%
\begin{figure*}
    \includegraphics[width=\linewidth]{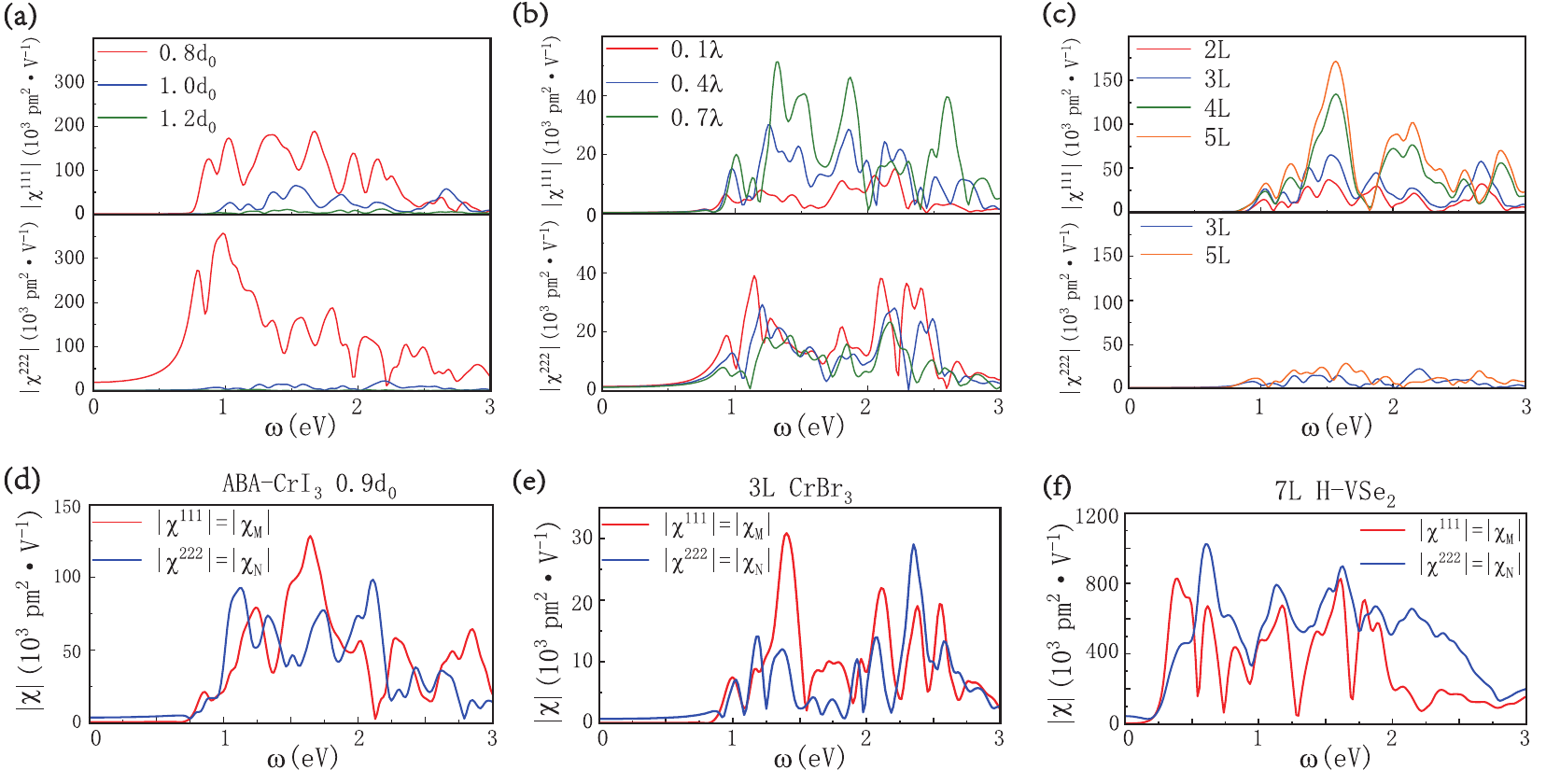}%[width=\linewidth] %[width=0.3\textwidth]
    \caption{Design principles of NLMO effects. The influence of interlayer spacing (a), SOC (b), and layer number (c) to $|\chi^{111}|$ which is the $\mathcal{T}$-odd term $\chi_{\rm M}$ and $|\chi^{222}|$ which is the $\mathcal{T}$-even term $\chi_{\rm N}$ in ABA-AFM CrI$_3$ multilayers. The SHG tensor components of ABA-AFM CrI$_3$ with interlayer distance 0.9$d_0$ (d), ABA-AFM CrBr$_3$ (e) and 7L H-VSe$_2$ (f). The $\mathcal{T}$-odd component $|\chi^{222}|$ in even-layer AFM CrI$_3$ are not shown.}  \label{Fig5}
\end{figure*}
%******************************************************************%

%Subtle magnetic orders detecting
The materials we investigated so far have several particular symmetry features. 
In general, NLMO effects are highly sensitive to changes in magnetic orders and can be used as a powerful tool to distinguish subtle magnetic states.
As an example, we considered a different magnetic order, $\uparrow\uparrow\downarrow$ for ABA stacking trilayer CrI$_3$, which we denoted it as the MIX state and its $\mathcal{T}$-pair state as tMIX.
ABA-MIX CrI$_3$ and ABA-AFM CrI$_3$ have the same net magnetic moment and tensor components as Tab. \ref{tab:sym} shows, \textcolor{black}{and their linear optical responses are similar as shown in Supplementary Note 4 \cite{SI},}
which make them difficult to be distinguished experimentally. 
However, in ABA-MIX CrI$_3$, each SHG tensor component contains both $\chi_{\rm N}$ and $\chi_{\rm M}$, and does not have a definite parity under $\mathcal{T}$, as shown in Fig. \ref{Fig4}(a). 
As a result, for LPL, the polarization-resolved SHG shows both rotation and magnitude change under $\mathcal{T}$ operation as shown at the bottom two plots in Fig. \ref{Fig4}(c). 
For CPL, the equivalence between helicity and magnetization reverse under $\eta_{\mathcal{T}}$ is breaking, as reflected by the remarkable difference between the upper and lower panels in the Fig. \ref{Fig4}(b).
Therefore, under the illumination of LPL (CPL) at $\omega_{\rm LPL}$ ($\omega_{\rm CPL}$) for ABA-AFM CrI$_3$, we can observe notable differences between ABA-AFM CrI$_3$ and ABA-MIX CrI$_3$, as shown in Fig. \ref{Fig4}(c). 

\textcolor{black}{In addition to the detection of full ordered magnetic ground states, the NLMO effects can also distinguish the influence from spin fluctuation at finite temperatures \cite{Tokura_FeGaO3_2004} and spin canting caused by an external field, as shown in Supplementary Figure 6 \cite{SI}.}

%--------------------------------------------------
 \subsection*{Controllable NLMO effects of representative 2D magnets}
%--------------------------------------------------
The giant NLMO effects manifest many intriguing applications as we have learned from above results, a necessary condition to achieve those effects is to have comparable $|\chi_{\rm N}|$ and $|\chi_{\rm M}|$.
However, in real materials, $|\chi_{\rm N}|$ and $|\chi_{\rm M}|$ may differ greatly which results in negligible NLMO effects.
In the following, taking ABA-AFM CrI$_3$ as a model system, we demonstrated several strategies to enhance NLMO effects, in which $\chi^{111}$ is the $\mathcal{T}$-odd term $\chi_{\rm M}$ and $\chi^{222}$ is the $\mathcal{T}$-even term $\chi_{\rm N}$ as summarized in Tab. \ref{tab:sym}

%diff d
The strength of interlayer interaction has great impact on properties of multilayer magnets \cite{Dixiao_2018_nanoletters_stacking,WeiRen_2020_VS2,Binghuang_2022_prb_SHG}, thus we firstly investigated the influence of interlayer distance $d$ in trilayer CrI$_3$, where $d_0$ is the interlayer distance of ABA-AFM CrI$_3$ at equilibrium.
As shown in Fig. \ref{Fig5}(a), both $|\chi_{\rm M}|$ and $|\chi_{\rm N}|$ change with $d$, but $|\chi_{\rm N}|$ is more sensitive to the variation of $d$. 
Therefore, a more comparable $|\chi_{\rm N}|$ and $|\chi_{\rm M}|$ can be realized by exerting hydrostatic pressure to multilayer CrI$_3$ along $z$-axis in experiments as shown in Fig. \ref{Fig5}(d).

%diff SOC
Next, we investigated the influence of spin-orbit coupling (SOC) to NLMO effects by artificially tuning the magnitude of SOC strength $\lambda$. 
As the Fig. \ref{Fig5}(b) shows, $\chi_{\rm M}$ is sensitive to the change of $\lambda$ while $\chi_{\rm N}$ changes slightly. 
As $|\chi_{\rm M}|$ is much larger than $|\chi_{\rm N}|$ in ABA-AFM CrI$_3$,
we replaced I by Br to reduce SOC and found ABA-AFM CrBr$_3$ posses very comparable $|\chi_{\rm N}|$ and $|\chi_{\rm M}|$ in a broad frequency range, as shown in Fig. \ref{Fig5}(e).

%diff layer-model
Furthermore, we found stacking and magnetic orders can have synergistic effect on engineering the relative and absolute value of $|\chi_{\rm N}|$ and $|\chi_{\rm M}|$.
As shown in Fig. \ref{Fig5}(c), in multilayer AB stacking AFM CrI$_3$, $\chi_{\rm N}$ is forbidden in even-layer structures due to the presence of $\mathcal{PT}$ symmetry and remains almost the same in odd-layer structures, while $\chi_{\rm M}$ increases with the layer number. 
The above observations can be simply understood as $\chi_{\rm N}$ from each layer alternates the sign with similar magnitude, while $\chi_{\rm M}$ from each layer has the same sign, due to the $\mathcal{PT}$ symmetry between the A and the B layers\cite{Heinz_SHG_2013}.
Inspired by this, as $|\chi_{\rm M}|$ is much smaller than $|\chi_{\rm N}|$ in ML H-VSe$_2$, we calculated the SHG susceptibility of 7-layer (7L) AA$^\prime$-AFM H-VSe$_2$, which also has $\mathcal{PT}$ symmetry between A and A$^\prime$ layers.
As expected, the peak value of $|\chi_{\rm M}|$ in 7-layer structure shown in Fig. \ref{Fig5}(f) is almost 7 times of that in ML H-VSe$_2$ (Fig. \ref{Fig2}f) while its $|\chi_{\rm N}|$ remains almost the same. 
$|\chi_{\rm N}|$ and $|\chi_{\rm M}|$ are comparable in 7L-AA$^\prime$ stacking H-VSe$_2$, indicating it can host giant NLMO effects. 
Similarly, by tuning the stacking sequence and magnetic order, we can achieve $\mathcal{P}$, $\mathcal{T}$ or neither symmetry between neighbouring layers and can realize an arbitrary control of the relative and absolute value of $|\chi_{\rm N}|$ and $|\chi_{\rm M}|$ (\textcolor{black}{Supplementary Figure 9}) \cite{SI}.

% %=================================
% \section{Summary}
% %==================================

To summarize, the simultaneous breaking of $\mathcal{P}$, $\mathcal{T}$ and their joint symmetry allows NLMO effects with the electric-dipole origin. In several representative 2D magnets with similar symmetry, we investigated and demonstrated NLMO angle $\theta_{\mathcal{T}}$ under LPL, NLMO intensity asymmetry $\eta_{\mathcal{T}}$ and SHG-CD $\eta_{\rm h}$ under CPL.
In particular, we discovered promising candidates with giant NLMO effects, including a near 90$^\circ$ polarization rotation and an on/off switching of certain helicity of SH light upon magnetization reversal, as well as $\pm 1$ SHG-CD within a certain magnetic configuration. 
These giant NLMO effects in candidate 2D magnets can be used not only to design atomically thin NLMO devices such as optical polarization switchers, switches and filters, but also to detect subtle magnetic orders in multilayer magnets such as ABA CrI$_3$. 
We further derived that the comparable magnitude of $|\chi_{\rm N}|$ and $|\chi_{\rm M}|$ is indispensable to achieve giant NLMO effects. 
Lastly and most importantly, we found the interlayer distance, magnitude of SOC, and the synergistic effect of stacking and magnetic orders could be used to control the relative and absolute magnitude of $|\chi_{\rm N}|$ and $|\chi_{\rm M}|$, which provides general design principles to achieve giant NLMO effects in 2D magnets. 
Our finding not only reveals several intriguing NLMO phenomena, but also pave the way to achieve subtle magnetization detection, giant and controllable NLMO effects in ultra-thin magneto-optical devices.

%=================================
\section*{Methods}\label{Methods}
%=================================
%--------------------------------------------------
 \subsection*{First-principles calculations}
%--------------------------------------------------
First-principles calculations were performed by Vienna $Ab \ initio$ Simulation Package \cite{VASP} with SOC included. The exchange-correlation functional was parameterized in the Perdew-Burke-Ernzerhof form \cite{PBE}, and the projector augmented-wave potential \cite{PAW} were used.
For the 3$d$ orbitals in magnetic ions Cr and V, the Hubbard $U$ of 3\,eV \cite{sivadas_CrI3_2018} and 1.16\,eV \cite{chungang-duan_CrI3_2018} were used. 
For layered materials, we used DFT-D3 form van der Waals correction without damping \cite{DFT-D3}.
The cut-off energy of plane waves was set to 450\,eV and 500\,eV for CrI$_3$-based materials and H-VSe$_2$-based materials, respectively.
The convergence criterion of force were set to 10\,meV/\AA\, and 1\,meV/\AA\ for CrI$_3$-based materials and H-VSe$_2$-based materials, respectively. 
Total energy is converged within $10^{-6}$\,eV.
$k$-point samplings of $13\times13\times1$ were used for CrI$_3$-based materials and $15\times15\times1$ for H-VSe$_2$-based materials. Vacuum thickness about 20 \AA\ was used in the calculations of 2D materials. %\WDZ{(ML H-VSe2 is 19.53A)}.
%--------------------------------------------------
 \subsection*{SHG calculations}
%--------------------------------------------------
%optics
After getting the converged electronic structures, we generated the maximally localized Wannier functions using Wannier90 \cite{wannier90} to build the tight-binding Hamiltonian and calculate optical responses \cite{Wang_NLO_Wannier}. 
\textcolor{black}{The $\mathcal{T}$ symmetry breaking is carefully treated in the calculations of SHG for magnets \cite{Chen_2022,Yan-2023-SHG_cal}.}
We obtained 56 maximally localized orbitals for each layer of CrI$_3$ or Cr$_2$I$_3$Br$_3$ and 22 for each layer of H-VSe$_2$.
The broadening factor of the Dirac delta function is taken to be 0.05\,eV. 
We found $50\times 50\times1$ and $150\times 150\times1$ kmesh samplings are enough for the converged SHG susceptibilities for CrI$_3$-based materials and H-VSe$_2$-based materials. \textcolor{black}{The band gap of trilayer CrI$_3$ has been scissored to 1.5 eV in the calculation of SHG susceptibility \cite{Yang_CrI3_2020}.}
% We use a $300\times 300\times1$ kmesh for the k-resolved SHG calculation.

%=================================
\section*{Data availability}
%=================================
The data that support the findings of this study are available from the corresponding author upon request.

\nocite{*}
% \bibliography{ref.bib}

%apsrev4-2.bst 2019-01-14 (MD) hand-edited version of apsrev4-1.bst
%Control: key (0)
%Control: author (8) initials jnrlst
%Control: editor formatted (1) identically to author
%Control: production of article title (0) allowed
%Control: page (0) single
%Control: year (1) truncated
%Control: production of eprint (0) enabled
\begin{thebibliography}{59}%
\makeatletter
\providecommand \@ifxundefined [1]{%
 \@ifx{#1\undefined}
}%
\providecommand \@ifnum [1]{%
 \ifnum #1\expandafter \@firstoftwo
 \else \expandafter \@secondoftwo
 \fi
}%
\providecommand \@ifx [1]{%
 \ifx #1\expandafter \@firstoftwo
 \else \expandafter \@secondoftwo
 \fi
}%
\providecommand \natexlab [1]{#1}%
\providecommand \enquote  [1]{``#1''}%
\providecommand \bibnamefont  [1]{#1}%
\providecommand \bibfnamefont [1]{#1}%
\providecommand \citenamefont [1]{#1}%
\providecommand \href@noop [0]{\@secondoftwo}%
\providecommand \href [0]{\begingroup \@sanitize@url \@href}%
\providecommand \@href[1]{\@@startlink{#1}\@@href}%
\providecommand \@@href[1]{\endgroup#1\@@endlink}%
\providecommand \@sanitize@url [0]{\catcode `\\12\catcode `\$12\catcode `\&12\catcode `\#12\catcode `\^12\catcode `\_12\catcode `\%12\relax}%
\providecommand \@@startlink[1]{}%
\providecommand \@@endlink[0]{}%
\providecommand \url  [0]{\begingroup\@sanitize@url \@url }%
\providecommand \@url [1]{\endgroup\@href {#1}{\urlprefix }}%
\providecommand \urlprefix  [0]{URL }%
\providecommand \Eprint [0]{\href }%
\providecommand \doibase [0]{https://doi.org/}%
\providecommand \selectlanguage [0]{\@gobble}%
\providecommand \bibinfo  [0]{\@secondoftwo}%
\providecommand \bibfield  [0]{\@secondoftwo}%
\providecommand \translation [1]{[#1]}%
\providecommand \BibitemOpen [0]{}%
\providecommand \bibitemStop [0]{}%
\providecommand \bibitemNoStop [0]{.\EOS\space}%
\providecommand \EOS [0]{\spacefactor3000\relax}%
\providecommand \BibitemShut  [1]{\csname bibitem#1\endcsname}%
\let\auto@bib@innerbib\@empty
%</preamble>
\bibitem [{\citenamefont {Sun}\ \emph {et~al.}(2019)\citenamefont {Sun} \emph {et~al.}}]{Wu_CrI3_2019}%
  \BibitemOpen
  \bibfield  {author} {\bibinfo {author} {\bibfnamefont {Z.}~\bibnamefont {Sun}} \emph {et~al.},\ }\bibfield  {title} {\bibinfo {title} {Giant nonreciprocal second-harmonic generation from antiferromagnetic bilayer {CrI$_3$}},\ }\href {https://doi.org/10.1038/s41586-019-1445-3} {\bibfield  {journal} {\bibinfo  {journal} {Nature}\ }\textbf {\bibinfo {volume} {572}},\ \bibinfo {pages} {497} (\bibinfo {year} {2019})}\BibitemShut {NoStop}%
\bibitem [{\citenamefont {Zhang}\ \emph {et~al.}(2019)\citenamefont {Zhang} \emph {et~al.}}]{Yan-2019-NC-MIC}%
  \BibitemOpen
  \bibfield  {author} {\bibinfo {author} {\bibfnamefont {Y.}~\bibnamefont {Zhang}} \emph {et~al.},\ }\bibfield  {title} {\bibinfo {title} {Switchable magnetic bulk photovoltaic effect in the two-dimensional magnet {CrI$_3$}},\ }\href@noop {} {\bibfield  {journal} {\bibinfo  {journal} {Nat. Commun.}\ }\textbf {\bibinfo {volume} {10}},\ \bibinfo {pages} {3783} (\bibinfo {year} {2019})}\BibitemShut {NoStop}%
\bibitem [{\citenamefont {Wang}\ and\ \citenamefont {Qian}(2020)}]{Qian-2020-npjCM-MIC}%
  \BibitemOpen
  \bibfield  {author} {\bibinfo {author} {\bibfnamefont {H.}~\bibnamefont {Wang}}\ and\ \bibinfo {author} {\bibfnamefont {X.}~\bibnamefont {Qian}},\ }\bibfield  {title} {\bibinfo {title} {Electrically and magnetically switchable nonlinear photocurrent in $\mathcal{PT}$-symmetric magnetic topological quantum materials},\ }\href@noop {} {\bibfield  {journal} {\bibinfo  {journal} {npj Comput. Mater.}\ }\textbf {\bibinfo {volume} {6}},\ \bibinfo {pages} {199} (\bibinfo {year} {2020})}\BibitemShut {NoStop}%
\bibitem [{\citenamefont {Fiebig}\ \emph {et~al.}(2000)\citenamefont {Fiebig} \emph {et~al.}}]{Pisarev_hRMnO3_2000}%
  \BibitemOpen
  \bibfield  {author} {\bibinfo {author} {\bibfnamefont {M.}~\bibnamefont {Fiebig}} \emph {et~al.},\ }\bibfield  {title} {\bibinfo {title} {Determination of the magnetic symmetry of hexagonal manganites by second harmonic generation},\ }\href {https://doi.org/10.1103/PhysRevLett.84.5620} {\bibfield  {journal} {\bibinfo  {journal} {Phys. Rev. Lett.}\ }\textbf {\bibinfo {volume} {84}},\ \bibinfo {pages} {5620} (\bibinfo {year} {2000})}\BibitemShut {NoStop}%
\bibitem [{\citenamefont {Fiebig}\ \emph {et~al.}(2005)\citenamefont {Fiebig}, \citenamefont {Pavlov},\ and\ \citenamefont {Pisarev}}]{Fiebig_rev_2005}%
  \BibitemOpen
  \bibfield  {author} {\bibinfo {author} {\bibfnamefont {M.}~\bibnamefont {Fiebig}}, \bibinfo {author} {\bibfnamefont {V.~V.}\ \bibnamefont {Pavlov}},\ and\ \bibinfo {author} {\bibfnamefont {R.~V.}\ \bibnamefont {Pisarev}},\ }\bibfield  {title} {\bibinfo {title} {Second-harmonic generation as a tool for studying electronic and magnetic structures of crystals: review},\ }\href {https://doi.org/10.1364/JOSAB.22.000096} {\bibfield  {journal} {\bibinfo  {journal} {J. Opt. Soc. Am. B}\ }\textbf {\bibinfo {volume} {22}},\ \bibinfo {pages} {96} (\bibinfo {year} {2005})}\BibitemShut {NoStop}%
\bibitem [{\citenamefont {Pavlov}\ \emph {et~al.}(1997)\citenamefont {Pavlov}, \citenamefont {Pisarev}, \citenamefont {Kirilyuk},\ and\ \citenamefont {Rasing}}]{Rasing_LPL_1997}%
  \BibitemOpen
  \bibfield  {author} {\bibinfo {author} {\bibfnamefont {V.~V.}\ \bibnamefont {Pavlov}}, \bibinfo {author} {\bibfnamefont {R.~V.}\ \bibnamefont {Pisarev}}, \bibinfo {author} {\bibfnamefont {A.}~\bibnamefont {Kirilyuk}},\ and\ \bibinfo {author} {\bibfnamefont {T.}~\bibnamefont {Rasing}},\ }\bibfield  {title} {\bibinfo {title} {Observation of a transversal nonlinear magneto-optical effect in thin magnetic garnet films},\ }\href {https://doi.org/10.1103/PhysRevLett.78.2004} {\bibfield  {journal} {\bibinfo  {journal} {Phys. Rev. Lett.}\ }\textbf {\bibinfo {volume} {78}},\ \bibinfo {pages} {2004} (\bibinfo {year} {1997})}\BibitemShut {NoStop}%
\bibitem [{\citenamefont {Straub}\ \emph {et~al.}(1996)\citenamefont {Straub}, \citenamefont {Vollmer},\ and\ \citenamefont {Kirschner}}]{Kirschner_Fe_1996}%
  \BibitemOpen
  \bibfield  {author} {\bibinfo {author} {\bibfnamefont {M.}~\bibnamefont {Straub}}, \bibinfo {author} {\bibfnamefont {R.}~\bibnamefont {Vollmer}},\ and\ \bibinfo {author} {\bibfnamefont {J.}~\bibnamefont {Kirschner}},\ }\bibfield  {title} {\bibinfo {title} {Surface magnetism of ultrathin $\mathit{\ensuremath{\gamma}}$-{Fe} films investigated by nonlinear magneto-optical {Kerr} effect},\ }\href {https://doi.org/10.1103/PhysRevLett.77.743} {\bibfield  {journal} {\bibinfo  {journal} {Phys. Rev. Lett.}\ }\textbf {\bibinfo {volume} {77}},\ \bibinfo {pages} {743} (\bibinfo {year} {1996})}\BibitemShut {NoStop}%
\bibitem [{\citenamefont {Kirilyuk}(2002)}]{Kirilyuk_rev_2002}%
  \BibitemOpen
  \bibfield  {author} {\bibinfo {author} {\bibfnamefont {A.}~\bibnamefont {Kirilyuk}},\ }\bibfield  {title} {\bibinfo {title} {Nonlinear optics in application to magnetic surfaces and thin films},\ }\href {https://doi.org/10.1088/0022-3727/35/21/202} {\bibfield  {journal} {\bibinfo  {journal} {J. Phys. D: Appl. Phys.}\ }\textbf {\bibinfo {volume} {35}},\ \bibinfo {pages} {R189} (\bibinfo {year} {2002})}\BibitemShut {NoStop}%
\bibitem [{\citenamefont {Huang}\ \emph {et~al.}(2017)\citenamefont {Huang} \emph {et~al.}}]{2D-magnetic-probing_2017_nature}%
  \BibitemOpen
  \bibfield  {author} {\bibinfo {author} {\bibfnamefont {B.}~\bibnamefont {Huang}} \emph {et~al.},\ }\bibfield  {title} {\bibinfo {title} {Layer-dependent ferromagnetism in a van der {Waals} crystal down to the monolayer limit},\ }\href@noop {} {\bibfield  {journal} {\bibinfo  {journal} {Nature}\ }\textbf {\bibinfo {volume} {546}},\ \bibinfo {pages} {270} (\bibinfo {year} {2017})}\BibitemShut {NoStop}%
\bibitem [{\citenamefont {Mak}\ \emph {et~al.}(2019)\citenamefont {Mak}, \citenamefont {Shan},\ and\ \citenamefont {Ralph}}]{Mak_2019}%
  \BibitemOpen
  \bibfield  {author} {\bibinfo {author} {\bibfnamefont {K.~F.}\ \bibnamefont {Mak}}, \bibinfo {author} {\bibfnamefont {J.}~\bibnamefont {Shan}},\ and\ \bibinfo {author} {\bibfnamefont {D.~C.}\ \bibnamefont {Ralph}},\ }\bibfield  {title} {\bibinfo {title} {Probing and controlling magnetic states in {2D} layered magnetic materials},\ }\href {https://doi.org/10.1038/s42254-019-0110-y} {\bibfield  {journal} {\bibinfo  {journal} {Nat. Rev. Phys.}\ }\textbf {\bibinfo {volume} {1}},\ \bibinfo {pages} {646} (\bibinfo {year} {2019})}\BibitemShut {NoStop}%
\bibitem [{\citenamefont {Wu}\ \emph {et~al.}(2023)\citenamefont {Wu} \emph {et~al.}}]{Wu-2023-ACSnano}%
  \BibitemOpen
  \bibfield  {author} {\bibinfo {author} {\bibfnamefont {S.}~\bibnamefont {Wu}} \emph {et~al.},\ }\bibfield  {title} {\bibinfo {title} {Extrinsic nonlinear {Kerr} rotation in topological materials under a magnetic field},\ }\href@noop {} {\bibfield  {journal} {\bibinfo  {journal} {ACS nano}\ }\textbf {\bibinfo {volume} {17}},\ \bibinfo {pages} {18905} (\bibinfo {year} {2023})}\BibitemShut {NoStop}%
\bibitem [{\citenamefont {Koopmans}\ \emph {et~al.}(1995)\citenamefont {Koopmans}, \citenamefont {Koerkamp}, \citenamefont {Rasing},\ and\ \citenamefont {van~den Berg}}]{Koopmans_LPL_1995}%
  \BibitemOpen
  \bibfield  {author} {\bibinfo {author} {\bibfnamefont {B.}~\bibnamefont {Koopmans}}, \bibinfo {author} {\bibfnamefont {M.~G.}\ \bibnamefont {Koerkamp}}, \bibinfo {author} {\bibfnamefont {T.}~\bibnamefont {Rasing}},\ and\ \bibinfo {author} {\bibfnamefont {H.}~\bibnamefont {van~den Berg}},\ }\bibfield  {title} {\bibinfo {title} {Observation of large {Kerr} angles in the nonlinear optical response from magnetic multilayers},\ }\href {https://doi.org/10.1103/PhysRevLett.74.3692} {\bibfield  {journal} {\bibinfo  {journal} {Phys. Rev. Lett.}\ }\textbf {\bibinfo {volume} {74}},\ \bibinfo {pages} {3692} (\bibinfo {year} {1995})}\BibitemShut {NoStop}%
\bibitem [{\citenamefont {Pershan}(1963)}]{Pershan_1963}%
  \BibitemOpen
  \bibfield  {author} {\bibinfo {author} {\bibfnamefont {P.~S.}\ \bibnamefont {Pershan}},\ }\bibfield  {title} {\bibinfo {title} {Nonlinear optical properties of solids: Energy considerations},\ }\href {https://doi.org/10.1103/PhysRev.130.919} {\bibfield  {journal} {\bibinfo  {journal} {Phys. Rev.}\ }\textbf {\bibinfo {volume} {130}},\ \bibinfo {pages} {919} (\bibinfo {year} {1963})}\BibitemShut {NoStop}%
\bibitem [{\citenamefont {Pan}\ \emph {et~al.}(1989)\citenamefont {Pan}, \citenamefont {Wei},\ and\ \citenamefont {Shen}}]{Shen_1989}%
  \BibitemOpen
  \bibfield  {author} {\bibinfo {author} {\bibfnamefont {R.-P.}\ \bibnamefont {Pan}}, \bibinfo {author} {\bibfnamefont {H.~D.}\ \bibnamefont {Wei}},\ and\ \bibinfo {author} {\bibfnamefont {Y.~R.}\ \bibnamefont {Shen}},\ }\bibfield  {title} {\bibinfo {title} {Optical second-harmonic generation from magnetized surfaces},\ }\href {https://doi.org/10.1103/PhysRevB.39.1229} {\bibfield  {journal} {\bibinfo  {journal} {Phys. Rev. B}\ }\textbf {\bibinfo {volume} {39}},\ \bibinfo {pages} {1229} (\bibinfo {year} {1989})}\BibitemShut {NoStop}%
\bibitem [{\citenamefont {Birss}(1964)}]{Birss_1964}%
  \BibitemOpen
  \bibfield  {author} {\bibinfo {author} {\bibfnamefont {R.~R.}\ \bibnamefont {Birss}},\ }\href@noop {} {\emph {\bibinfo {title} {Symmetry and magnetism}}},\ Vol.~\bibinfo {volume} {3}\ (\bibinfo  {publisher} {North-Holland Publishing Company},\ \bibinfo {year} {1964})\BibitemShut {NoStop}%
\bibitem [{\citenamefont {H\"ubner}\ and\ \citenamefont {Bennemann}(1989)}]{Hubner_1989}%
  \BibitemOpen
  \bibfield  {author} {\bibinfo {author} {\bibfnamefont {W.}~\bibnamefont {H\"ubner}}\ and\ \bibinfo {author} {\bibfnamefont {K.-H.}\ \bibnamefont {Bennemann}},\ }\bibfield  {title} {\bibinfo {title} {Nonlinear magneto-optical {Kerr} effect on a nickel surface},\ }\href {https://doi.org/10.1103/PhysRevB.40.5973} {\bibfield  {journal} {\bibinfo  {journal} {Phys. Rev. B}\ }\textbf {\bibinfo {volume} {40}},\ \bibinfo {pages} {5973} (\bibinfo {year} {1989})}\BibitemShut {NoStop}%
\bibitem [{\citenamefont {Shen}(1984)}]{shen1984principles}%
  \BibitemOpen
  \bibfield  {author} {\bibinfo {author} {\bibfnamefont {Y.-R.}\ \bibnamefont {Shen}},\ }\href@noop {} {\emph {\bibinfo {title} {Principles of nonlinear optics}}}\ (\bibinfo  {publisher} {Wiley-Interscience, New York, NY, USA},\ \bibinfo {year} {1984})\BibitemShut {NoStop}%
\bibitem [{SI()}]{SI}%
  \BibitemOpen
  \href@noop {} {\bibinfo {title} {Supporting information for ``giant and controllable nonlinear magneto-optical effects in two-dimensional magnets''}}\BibitemShut {NoStop}%
\bibitem [{\citenamefont {Reif}\ \emph {et~al.}(1991)\citenamefont {Reif}, \citenamefont {Zink}, \citenamefont {Schneider},\ and\ \citenamefont {Kirschner}}]{Kirschner_LPL_1991}%
  \BibitemOpen
  \bibfield  {author} {\bibinfo {author} {\bibfnamefont {J.}~\bibnamefont {Reif}}, \bibinfo {author} {\bibfnamefont {J.~C.}\ \bibnamefont {Zink}}, \bibinfo {author} {\bibfnamefont {C.-M.}\ \bibnamefont {Schneider}},\ and\ \bibinfo {author} {\bibfnamefont {J.}~\bibnamefont {Kirschner}},\ }\bibfield  {title} {\bibinfo {title} {Effects of surface magnetism on optical second harmonic generation},\ }\href {https://doi.org/10.1103/PhysRevLett.67.2878} {\bibfield  {journal} {\bibinfo  {journal} {Phys. Rev. Lett.}\ }\textbf {\bibinfo {volume} {67}},\ \bibinfo {pages} {2878} (\bibinfo {year} {1991})}\BibitemShut {NoStop}%
\bibitem [{\citenamefont {Kirilyuk}\ \emph {et~al.}(2000)\citenamefont {Kirilyuk}, \citenamefont {Pavlov}, \citenamefont {Pisarev},\ and\ \citenamefont {Rasing}}]{Kirilyuk_CPL_2000}%
  \BibitemOpen
  \bibfield  {author} {\bibinfo {author} {\bibfnamefont {A.}~\bibnamefont {Kirilyuk}}, \bibinfo {author} {\bibfnamefont {V.~V.}\ \bibnamefont {Pavlov}}, \bibinfo {author} {\bibfnamefont {R.~V.}\ \bibnamefont {Pisarev}},\ and\ \bibinfo {author} {\bibfnamefont {T.}~\bibnamefont {Rasing}},\ }\bibfield  {title} {\bibinfo {title} {Asymmetry of second harmonic generation in magnetic thin films under circular optical excitation},\ }\href {https://doi.org/10.1103/PhysRevB.61.R3796} {\bibfield  {journal} {\bibinfo  {journal} {Phys. Rev. B}\ }\textbf {\bibinfo {volume} {61}},\ \bibinfo {pages} {R3796} (\bibinfo {year} {2000})}\BibitemShut {NoStop}%
\bibitem [{\citenamefont {Gridnev}\ \emph {et~al.}(2001)\citenamefont {Gridnev}, \citenamefont {Pavlov}, \citenamefont {Pisarev}, \citenamefont {Kirilyuk},\ and\ \citenamefont {Rasing}}]{Kirilyuk_LPL_2001}%
  \BibitemOpen
  \bibfield  {author} {\bibinfo {author} {\bibfnamefont {V.~N.}\ \bibnamefont {Gridnev}}, \bibinfo {author} {\bibfnamefont {V.~V.}\ \bibnamefont {Pavlov}}, \bibinfo {author} {\bibfnamefont {R.~V.}\ \bibnamefont {Pisarev}}, \bibinfo {author} {\bibfnamefont {A.}~\bibnamefont {Kirilyuk}},\ and\ \bibinfo {author} {\bibfnamefont {T.}~\bibnamefont {Rasing}},\ }\bibfield  {title} {\bibinfo {title} {Second harmonic generation in anisotropic magnetic films},\ }\href {https://doi.org/10.1103/PhysRevB.63.184407} {\bibfield  {journal} {\bibinfo  {journal} {Phys. Rev. B}\ }\textbf {\bibinfo {volume} {63}},\ \bibinfo {pages} {184407} (\bibinfo {year} {2001})}\BibitemShut {NoStop}%
\bibitem [{\citenamefont {Train}\ \emph {et~al.}(2009)\citenamefont {Train}, \citenamefont {Nuida}, \citenamefont {Gheorghe}, \citenamefont {Gruselle},\ and\ \citenamefont {Ohkoshi}}]{Train_2009}%
  \BibitemOpen
  \bibfield  {author} {\bibinfo {author} {\bibfnamefont {C.}~\bibnamefont {Train}}, \bibinfo {author} {\bibfnamefont {T.}~\bibnamefont {Nuida}}, \bibinfo {author} {\bibfnamefont {R.}~\bibnamefont {Gheorghe}}, \bibinfo {author} {\bibfnamefont {M.}~\bibnamefont {Gruselle}},\ and\ \bibinfo {author} {\bibfnamefont {S.}~\bibnamefont {Ohkoshi}},\ }\bibfield  {title} {\bibinfo {title} {Large magnetization-induced second harmonic generation in an enantiopure chiral magnet},\ }\href {https://doi.org/10.1021/ja9061568} {\bibfield  {journal} {\bibinfo  {journal} {J. Am. Chem. Soc.}\ }\textbf {\bibinfo {volume} {131}},\ \bibinfo {pages} {16838} (\bibinfo {year} {2009})}\BibitemShut {NoStop}%
\bibitem [{\citenamefont {Ogawa}\ \emph {et~al.}(2004)\citenamefont {Ogawa}, \citenamefont {Kaneko}, \citenamefont {He}, \citenamefont {Yu}, \citenamefont {Arima},\ and\ \citenamefont {Tokura}}]{Tokura_FeGaO3_2004}%
  \BibitemOpen
  \bibfield  {author} {\bibinfo {author} {\bibfnamefont {Y.}~\bibnamefont {Ogawa}}, \bibinfo {author} {\bibfnamefont {Y.}~\bibnamefont {Kaneko}}, \bibinfo {author} {\bibfnamefont {J.~P.}\ \bibnamefont {He}}, \bibinfo {author} {\bibfnamefont {X.~Z.}\ \bibnamefont {Yu}}, \bibinfo {author} {\bibfnamefont {T.}~\bibnamefont {Arima}},\ and\ \bibinfo {author} {\bibfnamefont {Y.}~\bibnamefont {Tokura}},\ }\bibfield  {title} {\bibinfo {title} {Magnetization-induced second harmonic generation in a polar ferromagnet},\ }\href {https://doi.org/10.1103/PhysRevLett.92.047401} {\bibfield  {journal} {\bibinfo  {journal} {Phys. Rev. Lett.}\ }\textbf {\bibinfo {volume} {92}},\ \bibinfo {pages} {047401} (\bibinfo {year} {2004})}\BibitemShut {NoStop}%
\bibitem [{\citenamefont {Toyoda}\ \emph {et~al.}(2023)\citenamefont {Toyoda} \emph {et~al.}}]{Toyoda-2023-prm}%
  \BibitemOpen
  \bibfield  {author} {\bibinfo {author} {\bibfnamefont {S.}~\bibnamefont {Toyoda}} \emph {et~al.},\ }\bibfield  {title} {\bibinfo {title} {Magnetic-field switching of second-harmonic generation in noncentrosymmetric magnet {Eu$_2$MnSi$_2$O$_7$}},\ }\href@noop {} {\bibfield  {journal} {\bibinfo  {journal} {Phys. Rev. Mater.}\ }\textbf {\bibinfo {volume} {7}},\ \bibinfo {pages} {024403} (\bibinfo {year} {2023})}\BibitemShut {NoStop}%
\bibitem [{\citenamefont {Li}\ \emph {et~al.}(2013{\natexlab{a}})\citenamefont {Li} \emph {et~al.}}]{MoS2_2013_nanoletters}%
  \BibitemOpen
  \bibfield  {author} {\bibinfo {author} {\bibfnamefont {Y.}~\bibnamefont {Li}} \emph {et~al.},\ }\bibfield  {title} {\bibinfo {title} {Probing symmetry properties of few-layer {MoS$_2$} and {h-BN} by optical second-harmonic generation},\ }\href@noop {} {\bibfield  {journal} {\bibinfo  {journal} {Nano Lett.}\ }\textbf {\bibinfo {volume} {13}},\ \bibinfo {pages} {3329} (\bibinfo {year} {2013}{\natexlab{a}})}\BibitemShut {NoStop}%
\bibitem [{\citenamefont {Yin}\ \emph {et~al.}(2014)\citenamefont {Yin} \emph {et~al.}}]{MoS2_2014_Science}%
  \BibitemOpen
  \bibfield  {author} {\bibinfo {author} {\bibfnamefont {X.}~\bibnamefont {Yin}} \emph {et~al.},\ }\bibfield  {title} {\bibinfo {title} {Edge nonlinear optics on a {MoS2} atomic monolayer},\ }\href@noop {} {\bibfield  {journal} {\bibinfo  {journal} {Science}\ }\textbf {\bibinfo {volume} {344}},\ \bibinfo {pages} {488} (\bibinfo {year} {2014})}\BibitemShut {NoStop}%
\bibitem [{\citenamefont {Abdelwahab}\ \emph {et~al.}(2022)\citenamefont {Abdelwahab} \emph {et~al.}}]{NbOI2_2021_nature-photonics}%
  \BibitemOpen
  \bibfield  {author} {\bibinfo {author} {\bibfnamefont {I.}~\bibnamefont {Abdelwahab}} \emph {et~al.},\ }\bibfield  {title} {\bibinfo {title} {Giant second-harmonic generation in ferroelectric {NbOI$_2$}},\ }\href@noop {} {\bibfield  {journal} {\bibinfo  {journal} {Nat. Photonics}\ }\textbf {\bibinfo {volume} {16}},\ \bibinfo {pages} {644} (\bibinfo {year} {2022})}\BibitemShut {NoStop}%
\bibitem [{\citenamefont {Guo}\ \emph {et~al.}(2023)\citenamefont {Guo} \emph {et~al.}}]{NbOCl2_2023_nature}%
  \BibitemOpen
  \bibfield  {author} {\bibinfo {author} {\bibfnamefont {Q.}~\bibnamefont {Guo}} \emph {et~al.},\ }\bibfield  {title} {\bibinfo {title} {Ultrathin quantum light source with van der waals {NbOCl$_2$} crystal},\ }\href@noop {} {\bibfield  {journal} {\bibinfo  {journal} {Nature}\ }\textbf {\bibinfo {volume} {613}},\ \bibinfo {pages} {53} (\bibinfo {year} {2023})}\BibitemShut {NoStop}%
\bibitem [{\citenamefont {Kim}\ \emph {et~al.}(2013)\citenamefont {Kim} \emph {et~al.}}]{twisted-hBN_2013_nanoletters}%
  \BibitemOpen
  \bibfield  {author} {\bibinfo {author} {\bibfnamefont {C.-J.}\ \bibnamefont {Kim}} \emph {et~al.},\ }\bibfield  {title} {\bibinfo {title} {Stacking order dependent second harmonic generation and topological defects in {h-BN} bilayers},\ }\href@noop {} {\bibfield  {journal} {\bibinfo  {journal} {Nano Lett.}\ }\textbf {\bibinfo {volume} {13}},\ \bibinfo {pages} {5660} (\bibinfo {year} {2013})}\BibitemShut {NoStop}%
\bibitem [{\citenamefont {Yao}\ \emph {et~al.}(2021)\citenamefont {Yao} \emph {et~al.}}]{twist-hBN_2021_SA}%
  \BibitemOpen
  \bibfield  {author} {\bibinfo {author} {\bibfnamefont {K.}~\bibnamefont {Yao}} \emph {et~al.},\ }\bibfield  {title} {\bibinfo {title} {Enhanced tunable second harmonic generation from twistable interfaces and vertical superlattices in boron nitride homostructures},\ }\href@noop {} {\bibfield  {journal} {\bibinfo  {journal} {Sci. Adv.}\ }\textbf {\bibinfo {volume} {7}},\ \bibinfo {pages} {eabe8691} (\bibinfo {year} {2021})}\BibitemShut {NoStop}%
\bibitem [{\citenamefont {Chu}\ \emph {et~al.}(2020)\citenamefont {Chu} \emph {et~al.}}]{MnPS3_2020_prl}%
  \BibitemOpen
  \bibfield  {author} {\bibinfo {author} {\bibfnamefont {H.}~\bibnamefont {Chu}} \emph {et~al.},\ }\bibfield  {title} {\bibinfo {title} {Linear magnetoelectric phase in ultrathin {MnPS$_3$} probed by optical second harmonic generation},\ }\href@noop {} {\bibfield  {journal} {\bibinfo  {journal} {Phys. Rev. Lett}\ }\textbf {\bibinfo {volume} {124}},\ \bibinfo {pages} {027601} (\bibinfo {year} {2020})}\BibitemShut {NoStop}%
\bibitem [{\citenamefont {Lee}\ \emph {et~al.}(2021)\citenamefont {Lee} \emph {et~al.}}]{CrSBr_2021_nanoletters}%
  \BibitemOpen
  \bibfield  {author} {\bibinfo {author} {\bibfnamefont {K.}~\bibnamefont {Lee}} \emph {et~al.},\ }\bibfield  {title} {\bibinfo {title} {Magnetic order and symmetry in the {2D} semiconductor {CrSBr}},\ }\href@noop {} {\bibfield  {journal} {\bibinfo  {journal} {Nano Lett.}\ }\textbf {\bibinfo {volume} {21}},\ \bibinfo {pages} {3511} (\bibinfo {year} {2021})}\BibitemShut {NoStop}%
\bibitem [{\citenamefont {Fonseca}\ \emph {et~al.}(2022)\citenamefont {Fonseca} \emph {et~al.}}]{Xu_MBT_2022}%
  \BibitemOpen
  \bibfield  {author} {\bibinfo {author} {\bibfnamefont {J.}~\bibnamefont {Fonseca}} \emph {et~al.},\ }\bibfield  {title} {\bibinfo {title} {Anomalous second harmonic generation from atomically thin {MnBi$_2$Te$_4$}},\ }\href {https://doi.org/10.1021/acs.nanolett.2c04010} {\bibfield  {journal} {\bibinfo  {journal} {Nano Lett.}\ }\textbf {\bibinfo {volume} {22}},\ \bibinfo {pages} {10134} (\bibinfo {year} {2022})}\BibitemShut {NoStop}%
\bibitem [{\citenamefont {Song}\ \emph {et~al.}(2022)\citenamefont {Song} \emph {et~al.}}]{NiI2_2022_Nature}%
  \BibitemOpen
  \bibfield  {author} {\bibinfo {author} {\bibfnamefont {Q.}~\bibnamefont {Song}} \emph {et~al.},\ }\bibfield  {title} {\bibinfo {title} {Evidence for a single-layer van der waals multiferroic},\ }\href@noop {} {\bibfield  {journal} {\bibinfo  {journal} {Nature}\ }\textbf {\bibinfo {volume} {602}},\ \bibinfo {pages} {601} (\bibinfo {year} {2022})}\BibitemShut {NoStop}%
\bibitem [{\citenamefont {Wang}\ \emph {et~al.}(2017)\citenamefont {Wang}, \citenamefont {Liu}, \citenamefont {Kang}, \citenamefont {Gu}, \citenamefont {Xu},\ and\ \citenamefont {Duan}}]{Wang_NLO_Wannier}%
  \BibitemOpen
  \bibfield  {author} {\bibinfo {author} {\bibfnamefont {C.}~\bibnamefont {Wang}}, \bibinfo {author} {\bibfnamefont {X.}~\bibnamefont {Liu}}, \bibinfo {author} {\bibfnamefont {L.}~\bibnamefont {Kang}}, \bibinfo {author} {\bibfnamefont {B.-L.}\ \bibnamefont {Gu}}, \bibinfo {author} {\bibfnamefont {Y.}~\bibnamefont {Xu}},\ and\ \bibinfo {author} {\bibfnamefont {W.}~\bibnamefont {Duan}},\ }\bibfield  {title} {\bibinfo {title} {First-principles calculation of nonlinear optical responses by {Wannier} interpolation},\ }\href {https://doi.org/10.1103/PhysRevB.96.115147} {\bibfield  {journal} {\bibinfo  {journal} {Phys. Rev. B}\ }\textbf {\bibinfo {volume} {96}},\ \bibinfo {pages} {115147} (\bibinfo {year} {2017})}\BibitemShut {NoStop}%
\bibitem [{\citenamefont {Chen}\ \emph {et~al.}(2022)\citenamefont {Chen}, \citenamefont {Ye}, \citenamefont {Zou}, \citenamefont {Gu}, \citenamefont {Xu},\ and\ \citenamefont {Duan}}]{Chen_2022}%
  \BibitemOpen
  \bibfield  {author} {\bibinfo {author} {\bibfnamefont {H.}~\bibnamefont {Chen}}, \bibinfo {author} {\bibfnamefont {M.}~\bibnamefont {Ye}}, \bibinfo {author} {\bibfnamefont {N.}~\bibnamefont {Zou}}, \bibinfo {author} {\bibfnamefont {B.-L.}\ \bibnamefont {Gu}}, \bibinfo {author} {\bibfnamefont {Y.}~\bibnamefont {Xu}},\ and\ \bibinfo {author} {\bibfnamefont {W.}~\bibnamefont {Duan}},\ }\bibfield  {title} {\bibinfo {title} {Basic formulation and first-principles implementation of nonlinear magneto-optical effects},\ }\href {https://doi.org/10.1103/PhysRevB.105.075123} {\bibfield  {journal} {\bibinfo  {journal} {Phys. Rev. B}\ }\textbf {\bibinfo {volume} {105}},\ \bibinfo {pages} {075123} (\bibinfo {year} {2022})}\BibitemShut {NoStop}%
\bibitem [{\citenamefont {McGuire}\ \emph {et~al.}(2015)\citenamefont {McGuire}, \citenamefont {Dixit}, \citenamefont {Cooper},\ and\ \citenamefont {Sales}}]{mcguire_CrI3_2015}%
  \BibitemOpen
  \bibfield  {author} {\bibinfo {author} {\bibfnamefont {M.~A.}\ \bibnamefont {McGuire}}, \bibinfo {author} {\bibfnamefont {H.}~\bibnamefont {Dixit}}, \bibinfo {author} {\bibfnamefont {V.~R.}\ \bibnamefont {Cooper}},\ and\ \bibinfo {author} {\bibfnamefont {B.~C.}\ \bibnamefont {Sales}},\ }\bibfield  {title} {\bibinfo {title} {Coupling of crystal structure and magnetism in the layered, ferromagnetic insulator {CrI$_3$}},\ }\href@noop {} {\bibfield  {journal} {\bibinfo  {journal} {Chem. Mater.}\ }\textbf {\bibinfo {volume} {27}},\ \bibinfo {pages} {612} (\bibinfo {year} {2015})}\BibitemShut {NoStop}%
\bibitem [{\citenamefont {Sivadas}\ \emph {et~al.}(2018{\natexlab{a}})\citenamefont {Sivadas}, \citenamefont {Okamoto}, \citenamefont {Xu}, \citenamefont {Fennie},\ and\ \citenamefont {Xiao}}]{sivadas_CrI3_2018}%
  \BibitemOpen
  \bibfield  {author} {\bibinfo {author} {\bibfnamefont {N.}~\bibnamefont {Sivadas}}, \bibinfo {author} {\bibfnamefont {S.}~\bibnamefont {Okamoto}}, \bibinfo {author} {\bibfnamefont {X.}~\bibnamefont {Xu}}, \bibinfo {author} {\bibfnamefont {C.~J.}\ \bibnamefont {Fennie}},\ and\ \bibinfo {author} {\bibfnamefont {D.}~\bibnamefont {Xiao}},\ }\bibfield  {title} {\bibinfo {title} {Stacking-dependent magnetism in bilayer {CrI$_3$}},\ }\href@noop {} {\bibfield  {journal} {\bibinfo  {journal} {Nano Lett.}\ }\textbf {\bibinfo {volume} {18}},\ \bibinfo {pages} {7658} (\bibinfo {year} {2018}{\natexlab{a}})}\BibitemShut {NoStop}%
\bibitem [{\citenamefont {Jiang}\ \emph {et~al.}(2019)\citenamefont {Jiang} \emph {et~al.}}]{jiang_CrI3_2019}%
  \BibitemOpen
  \bibfield  {author} {\bibinfo {author} {\bibfnamefont {P.}~\bibnamefont {Jiang}} \emph {et~al.},\ }\bibfield  {title} {\bibinfo {title} {Stacking tunable interlayer magnetism in bilayer {CrI$_3$}},\ }\href@noop {} {\bibfield  {journal} {\bibinfo  {journal} {Phys. Rev. B}\ }\textbf {\bibinfo {volume} {99}},\ \bibinfo {pages} {144401} (\bibinfo {year} {2019})}\BibitemShut {NoStop}%
\bibitem [{\citenamefont {Jang}\ \emph {et~al.}(2019)\citenamefont {Jang}, \citenamefont {Jeong}, \citenamefont {Yoon}, \citenamefont {Ryee},\ and\ \citenamefont {Han}}]{jang_CrI3_2019_microscopic}%
  \BibitemOpen
  \bibfield  {author} {\bibinfo {author} {\bibfnamefont {S.~W.}\ \bibnamefont {Jang}}, \bibinfo {author} {\bibfnamefont {M.~Y.}\ \bibnamefont {Jeong}}, \bibinfo {author} {\bibfnamefont {H.}~\bibnamefont {Yoon}}, \bibinfo {author} {\bibfnamefont {S.}~\bibnamefont {Ryee}},\ and\ \bibinfo {author} {\bibfnamefont {M.~J.}\ \bibnamefont {Han}},\ }\bibfield  {title} {\bibinfo {title} {Microscopic understanding of magnetic interactions in bilayer {CrI$_3$}},\ }\href@noop {} {\bibfield  {journal} {\bibinfo  {journal} {Phys. Rev. Mater.}\ }\textbf {\bibinfo {volume} {3}},\ \bibinfo {pages} {031001} (\bibinfo {year} {2019})}\BibitemShut {NoStop}%
\bibitem [{\citenamefont {Soriano}\ \emph {et~al.}(2019)\citenamefont {Soriano}, \citenamefont {Cardoso},\ and\ \citenamefont {Fern{\'a}ndez-Rossier}}]{soriano_CrI3_2019}%
  \BibitemOpen
  \bibfield  {author} {\bibinfo {author} {\bibfnamefont {D.}~\bibnamefont {Soriano}}, \bibinfo {author} {\bibfnamefont {C.}~\bibnamefont {Cardoso}},\ and\ \bibinfo {author} {\bibfnamefont {J.}~\bibnamefont {Fern{\'a}ndez-Rossier}},\ }\bibfield  {title} {\bibinfo {title} {Interplay between interlayer exchange and stacking in {CrI$_3$} bilayers},\ }\href@noop {} {\bibfield  {journal} {\bibinfo  {journal} {Solid State Commun.}\ }\textbf {\bibinfo {volume} {299}},\ \bibinfo {pages} {113662} (\bibinfo {year} {2019})}\BibitemShut {NoStop}%
\bibitem [{\citenamefont {Bonilla}\ \emph {et~al.}(2018)\citenamefont {Bonilla} \emph {et~al.}}]{bonilla_VSe2_2018}%
  \BibitemOpen
  \bibfield  {author} {\bibinfo {author} {\bibfnamefont {M.}~\bibnamefont {Bonilla}} \emph {et~al.},\ }\bibfield  {title} {\bibinfo {title} {Strong room-temperature ferromagnetism in {VSe$_2$} monolayers on van der waals substrates},\ }\href@noop {} {\bibfield  {journal} {\bibinfo  {journal} {Nat. Nanotechnol.}\ }\textbf {\bibinfo {volume} {13}},\ \bibinfo {pages} {289} (\bibinfo {year} {2018})}\BibitemShut {NoStop}%
\bibitem [{\citenamefont {Wang}\ and\ \citenamefont {others}()\citenamefont {Wang} \emph {et~al.}}]{wang_VSe2_2021}%
  \BibitemOpen
  \bibfield  {author} {\bibinfo {author} {\bibfnamefont {X.}~\bibnamefont {Wang}} \emph {et~al.},\ }\bibfield  {title} {\bibinfo {title} {Ferromagnetism in {2D} vanadium diselenide},\ }\href@noop {} {\ }\BibitemShut {NoStop}%
\bibitem [{\citenamefont {Song}\ \emph {et~al.}(2020)\citenamefont {Song}, \citenamefont {Fei}, \citenamefont {Zhu},\ and\ \citenamefont {Yang}}]{Yang_CrI3_2020}%
  \BibitemOpen
  \bibfield  {author} {\bibinfo {author} {\bibfnamefont {W.}~\bibnamefont {Song}}, \bibinfo {author} {\bibfnamefont {R.}~\bibnamefont {Fei}}, \bibinfo {author} {\bibfnamefont {L.}~\bibnamefont {Zhu}},\ and\ \bibinfo {author} {\bibfnamefont {L.}~\bibnamefont {Yang}},\ }\bibfield  {title} {\bibinfo {title} {Nonreciprocal second-harmonic generation in few-layer chromium triiodide},\ }\href {https://doi.org/10.1103/PhysRevB.102.045411} {\bibfield  {journal} {\bibinfo  {journal} {Phys. Rev. B}\ }\textbf {\bibinfo {volume} {102}},\ \bibinfo {pages} {045411} (\bibinfo {year} {2020})}\BibitemShut {NoStop}%
\bibitem [{\citenamefont {Petralli-Mallow}\ \emph {et~al.}(1993)\citenamefont {Petralli-Mallow}, \citenamefont {Wong}, \citenamefont {Byers}, \citenamefont {Yee},\ and\ \citenamefont {Hicks}}]{Hicks_SHGCD_1993}%
  \BibitemOpen
  \bibfield  {author} {\bibinfo {author} {\bibfnamefont {T.}~\bibnamefont {Petralli-Mallow}}, \bibinfo {author} {\bibfnamefont {T.~M.}\ \bibnamefont {Wong}}, \bibinfo {author} {\bibfnamefont {J.~D.}\ \bibnamefont {Byers}}, \bibinfo {author} {\bibfnamefont {H.~I.}\ \bibnamefont {Yee}},\ and\ \bibinfo {author} {\bibfnamefont {J.~M.}\ \bibnamefont {Hicks}},\ }\bibfield  {title} {\bibinfo {title} {Circular dichroism spectroscopy at interfaces: a surface second harmonic generation study},\ }\href {https://doi.org/10.1021/j100109a022} {\bibfield  {journal} {\bibinfo  {journal} {J. Phys. Chem. C.}\ }\textbf {\bibinfo {volume} {97}},\ \bibinfo {pages} {1383} (\bibinfo {year} {1993})}\BibitemShut {NoStop}%
\bibitem [{\citenamefont {Byers}\ \emph {et~al.}(1994)\citenamefont {Byers}, \citenamefont {Yee}, \citenamefont {Petralli-Mallow},\ and\ \citenamefont {Hicks}}]{Hicks_SHGCD_1994}%
  \BibitemOpen
  \bibfield  {author} {\bibinfo {author} {\bibfnamefont {J.~D.}\ \bibnamefont {Byers}}, \bibinfo {author} {\bibfnamefont {H.~I.}\ \bibnamefont {Yee}}, \bibinfo {author} {\bibfnamefont {T.}~\bibnamefont {Petralli-Mallow}},\ and\ \bibinfo {author} {\bibfnamefont {J.~M.}\ \bibnamefont {Hicks}},\ }\bibfield  {title} {\bibinfo {title} {Second-harmonic generation circular-dichroism spectroscopy from chiral monolayers},\ }\href {https://doi.org/10.1103/PhysRevB.49.14643} {\bibfield  {journal} {\bibinfo  {journal} {Phys. Rev. B}\ }\textbf {\bibinfo {volume} {49}},\ \bibinfo {pages} {14643} (\bibinfo {year} {1994})}\BibitemShut {NoStop}%
\bibitem [{\citenamefont {Rodrigues}\ \emph {et~al.}()\citenamefont {Rodrigues}, \citenamefont {Lan}, \citenamefont {Kang}, \citenamefont {Cui},\ and\ \citenamefont {Cai}}]{Cai_SHGCD_2014}%
  \BibitemOpen
  \bibfield  {author} {\bibinfo {author} {\bibfnamefont {S.~P.}\ \bibnamefont {Rodrigues}}, \bibinfo {author} {\bibfnamefont {S.}~\bibnamefont {Lan}}, \bibinfo {author} {\bibfnamefont {L.}~\bibnamefont {Kang}}, \bibinfo {author} {\bibfnamefont {Y.}~\bibnamefont {Cui}},\ and\ \bibinfo {author} {\bibfnamefont {W.}~\bibnamefont {Cai}},\ }\bibfield  {title} {\bibinfo {title} {Nonlinear imaging and spectroscopy of chiral metamaterials},\ }\href {https://doi.org/https://doi.org/10.1002/adma.201402293} {\bibfield  {journal} {\bibinfo  {journal} {Adv. Mater.}\ }\textbf {\bibinfo {volume} {26}},\ \bibinfo {pages} {6157}}\BibitemShut {NoStop}%
\bibitem [{\citenamefont {Zu}\ \emph {et~al.}(2022)\citenamefont {Zu} \emph {et~al.}}]{Chen-2022-npjCM-SHG_computation}%
  \BibitemOpen
  \bibfield  {author} {\bibinfo {author} {\bibfnamefont {R.}~\bibnamefont {Zu}} \emph {et~al.},\ }\bibfield  {title} {\bibinfo {title} {Analytical and numerical modeling of optical second harmonic generation in anisotropic crystals using {♯SHAARP} package},\ }\href@noop {} {\bibfield  {journal} {\bibinfo  {journal} {npj Comput. Mater.}\ }\textbf {\bibinfo {volume} {8}},\ \bibinfo {pages} {246} (\bibinfo {year} {2022})}\BibitemShut {NoStop}%
\bibitem [{\citenamefont {Sivadas}\ \emph {et~al.}(2018{\natexlab{b}})\citenamefont {Sivadas}, \citenamefont {Okamoto}, \citenamefont {Xu}, \citenamefont {Fennie},\ and\ \citenamefont {Xiao}}]{Dixiao_2018_nanoletters_stacking}%
  \BibitemOpen
  \bibfield  {author} {\bibinfo {author} {\bibfnamefont {N.}~\bibnamefont {Sivadas}}, \bibinfo {author} {\bibfnamefont {S.}~\bibnamefont {Okamoto}}, \bibinfo {author} {\bibfnamefont {X.}~\bibnamefont {Xu}}, \bibinfo {author} {\bibfnamefont {C.~J.}\ \bibnamefont {Fennie}},\ and\ \bibinfo {author} {\bibfnamefont {D.}~\bibnamefont {Xiao}},\ }\bibfield  {title} {\bibinfo {title} {Stacking-dependent magnetism in bilayer {CrI$_3$}},\ }\href@noop {} {\bibfield  {journal} {\bibinfo  {journal} {Nano Lett.}\ }\textbf {\bibinfo {volume} {18}},\ \bibinfo {pages} {7658} (\bibinfo {year} {2018}{\natexlab{b}})}\BibitemShut {NoStop}%
\bibitem [{\citenamefont {Liu}\ \emph {et~al.}(2020)\citenamefont {Liu}, \citenamefont {Pyatakov},\ and\ \citenamefont {Ren}}]{WeiRen_2020_VS2}%
  \BibitemOpen
  \bibfield  {author} {\bibinfo {author} {\bibfnamefont {X.}~\bibnamefont {Liu}}, \bibinfo {author} {\bibfnamefont {A.~P.}\ \bibnamefont {Pyatakov}},\ and\ \bibinfo {author} {\bibfnamefont {W.}~\bibnamefont {Ren}},\ }\bibfield  {title} {\bibinfo {title} {Magnetoelectric coupling in multiferroic bilayer {VS$_2$}},\ }\href@noop {} {\bibfield  {journal} {\bibinfo  {journal} {Phys. Rev. Lett.}\ }\textbf {\bibinfo {volume} {125}},\ \bibinfo {pages} {247601} (\bibinfo {year} {2020})}\BibitemShut {NoStop}%
\bibitem [{\citenamefont {Jiang}\ \emph {et~al.}(2022)\citenamefont {Jiang}, \citenamefont {Kang},\ and\ \citenamefont {Huang}}]{Binghuang_2022_prb_SHG}%
  \BibitemOpen
  \bibfield  {author} {\bibinfo {author} {\bibfnamefont {X.}~\bibnamefont {Jiang}}, \bibinfo {author} {\bibfnamefont {L.}~\bibnamefont {Kang}},\ and\ \bibinfo {author} {\bibfnamefont {B.}~\bibnamefont {Huang}},\ }\bibfield  {title} {\bibinfo {title} {Role of interlayer coupling in second harmonic generation in bilayer transition metal dichalcogenides},\ }\href@noop {} {\bibfield  {journal} {\bibinfo  {journal} {Phys. Rev. B}\ }\textbf {\bibinfo {volume} {105}},\ \bibinfo {pages} {045415} (\bibinfo {year} {2022})}\BibitemShut {NoStop}%
\bibitem [{\citenamefont {Li}\ \emph {et~al.}(2013{\natexlab{b}})\citenamefont {Li} \emph {et~al.}}]{Heinz_SHG_2013}%
  \BibitemOpen
  \bibfield  {author} {\bibinfo {author} {\bibfnamefont {Y.}~\bibnamefont {Li}} \emph {et~al.},\ }\bibfield  {title} {\bibinfo {title} {Probing symmetry properties of few-layer {MoS$_2$} and {h-BN} by optical second-harmonic generation},\ }\href {https://doi.org/10.1021/nl401561r} {\bibfield  {journal} {\bibinfo  {journal} {Nano Lett.}\ }\textbf {\bibinfo {volume} {13}},\ \bibinfo {pages} {3329} (\bibinfo {year} {2013}{\natexlab{b}})}\BibitemShut {NoStop}%
\bibitem [{\citenamefont {Kresse}\ and\ \citenamefont {Furthm\"uller}(1996)}]{VASP}%
  \BibitemOpen
  \bibfield  {author} {\bibinfo {author} {\bibfnamefont {G.}~\bibnamefont {Kresse}}\ and\ \bibinfo {author} {\bibfnamefont {J.}~\bibnamefont {Furthm\"uller}},\ }\bibfield  {title} {\bibinfo {title} {Efficient iterative schemes for ab initio total-energy calculations using a plane-wave basis set},\ }\href {https://doi.org/10.1103/PhysRevB.54.11169} {\bibfield  {journal} {\bibinfo  {journal} {Phys. Rev. B}\ }\textbf {\bibinfo {volume} {54}},\ \bibinfo {pages} {11169} (\bibinfo {year} {1996})}\BibitemShut {NoStop}%
\bibitem [{\citenamefont {Perdew}\ \emph {et~al.}(1996)\citenamefont {Perdew}, \citenamefont {Burke},\ and\ \citenamefont {Ernzerhof}}]{PBE}%
  \BibitemOpen
  \bibfield  {author} {\bibinfo {author} {\bibfnamefont {J.~P.}\ \bibnamefont {Perdew}}, \bibinfo {author} {\bibfnamefont {K.}~\bibnamefont {Burke}},\ and\ \bibinfo {author} {\bibfnamefont {M.}~\bibnamefont {Ernzerhof}},\ }\bibfield  {title} {\bibinfo {title} {Generalized gradient approximation made simple},\ }\href {https://doi.org/10.1103/PhysRevLett.77.3865} {\bibfield  {journal} {\bibinfo  {journal} {Phys. Rev. Lett.}\ }\textbf {\bibinfo {volume} {77}},\ \bibinfo {pages} {3865} (\bibinfo {year} {1996})}\BibitemShut {NoStop}%
\bibitem [{\citenamefont {Kresse}\ and\ \citenamefont {Joubert}(1999)}]{PAW}%
  \BibitemOpen
  \bibfield  {author} {\bibinfo {author} {\bibfnamefont {G.}~\bibnamefont {Kresse}}\ and\ \bibinfo {author} {\bibfnamefont {D.}~\bibnamefont {Joubert}},\ }\bibfield  {title} {\bibinfo {title} {From ultrasoft pseudopotentials to the projector augmented-wave method},\ }\href {https://doi.org/10.1103/PhysRevB.59.1758} {\bibfield  {journal} {\bibinfo  {journal} {Phys. Rev. B}\ }\textbf {\bibinfo {volume} {59}},\ \bibinfo {pages} {1758} (\bibinfo {year} {1999})}\BibitemShut {NoStop}%
\bibitem [{\citenamefont {Gong}\ \emph {et~al.}(2018)\citenamefont {Gong} \emph {et~al.}}]{chungang-duan_CrI3_2018}%
  \BibitemOpen
  \bibfield  {author} {\bibinfo {author} {\bibfnamefont {S.-J.}\ \bibnamefont {Gong}} \emph {et~al.},\ }\bibfield  {title} {\bibinfo {title} {Electrically induced {2D} half-metallic antiferromagnets and spin field effect transistors},\ }\href@noop {} {\bibfield  {journal} {\bibinfo  {journal} {Proc. Natl. Acad. Sci.}\ }\textbf {\bibinfo {volume} {115}},\ \bibinfo {pages} {8511} (\bibinfo {year} {2018})}\BibitemShut {NoStop}%
\bibitem [{\citenamefont {Grimme}\ \emph {et~al.}(2010)\citenamefont {Grimme}, \citenamefont {Antony}, \citenamefont {Ehrlich},\ and\ \citenamefont {Krieg}}]{DFT-D3}%
  \BibitemOpen
  \bibfield  {author} {\bibinfo {author} {\bibfnamefont {S.}~\bibnamefont {Grimme}}, \bibinfo {author} {\bibfnamefont {J.}~\bibnamefont {Antony}}, \bibinfo {author} {\bibfnamefont {S.}~\bibnamefont {Ehrlich}},\ and\ \bibinfo {author} {\bibfnamefont {H.}~\bibnamefont {Krieg}},\ }\bibfield  {title} {\bibinfo {title} {A consistent and accurate ab initio parametrization of density functional dispersion correction {(DFT-D)} for the 94 elements {H-Pu}},\ }\href {https://doi.org/https://doi.org/10.1063/1.3382344} {\bibfield  {journal} {\bibinfo  {journal} {J. Chem. Phys.}\ }\textbf {\bibinfo {volume} {132}},\ \bibinfo {pages} {154104} (\bibinfo {year} {2010})}\BibitemShut {NoStop}%
\bibitem [{\citenamefont {Mostofi}\ \emph {et~al.}(2014)\citenamefont {Mostofi} \emph {et~al.}}]{wannier90}%
  \BibitemOpen
  \bibfield  {author} {\bibinfo {author} {\bibfnamefont {A.~A.}\ \bibnamefont {Mostofi}} \emph {et~al.},\ }\bibfield  {title} {\bibinfo {title} {An updated version of {Wannier90}: A tool for obtaining maximally-localised {Wannier} functions},\ }\href {https://doi.org/https://doi.org/10.1016/j.cpc.2014.05.003} {\bibfield  {journal} {\bibinfo  {journal} {Comput. Phys. Commun.}\ }\textbf {\bibinfo {volume} {185}},\ \bibinfo {pages} {2309} (\bibinfo {year} {2014})}\BibitemShut {NoStop}%
\bibitem [{\citenamefont {Kaplan}\ \emph {et~al.}(2023)\citenamefont {Kaplan} \emph {et~al.}}]{Yan-2023-SHG_cal}%
  \BibitemOpen
  \bibfield  {author} {\bibinfo {author} {\bibfnamefont {D.}~\bibnamefont {Kaplan}} \emph {et~al.},\ }\bibfield  {title} {\bibinfo {title} {Unifying semiclassics and quantum perturbation theory at nonlinear order},\ }\href@noop {} {\bibfield  {journal} {\bibinfo  {journal} {SciPost Phys.}\ }\textbf {\bibinfo {volume} {14}},\ \bibinfo {pages} {082} (\bibinfo {year} {2023})}\BibitemShut {NoStop}%
\end{thebibliography}%
%

%=================================
\section*{Acknowledgements}
%=================================
We thank Lu Wang for helpful discussions about plotting. This work was supported by the Basic Science Center Project of NSFC (Grant No.52388201), the National Science Fund for Distinguished Young Scholars (Grant No.12025405), the Beijing Advanced Innovation Center for Future Chip (ICFC), the Beijing Advanced Innovation Center for Materials Genome Engineering, and NSAF (Grant No.U2330401).
%=================================
\section*{Competing interests}
%=================================
The authors declare no competing interests.
%=================================
\section*{AUTHOR CONTRIBUTIONS}
%=================================
M.Y. conceived the project. D.W. carried out the DFT and NLMO calculations, and analyzed the data. D.W. and M.Y. performed the analytical derivation of NLMO effects and analyzed the results. D.W. and M.Y. wrote the manuscript in consultation with all the authors. All authors discussed the results. M.Y. and Y.X. supervised the project. 
%=================================
\section*{Additional information}
%=================================
\textbf{Supplementary Information} The online version contains supplementary material available at https://doi.org/xxx/xxx.\\
\textbf{Correspondence} and requests for materials should be addressed to M.Y. and Y.X.\\
\textbf{Reprints and permission information} is available at https://xxx/xxx.
\end{document}